\documentclass[10pt,letterpaper]{article}
\usepackage{times}
\usepackage{amssymb,amsmath,graphicx}

\begin{document}

\title{Optimal pupil apodizations for arbitrary apertures}

\author{A. Carlotti, R. Vanderbei and N. J. Kasdin}

\maketitle





\begin{abstract} We present here fully optimized two-dimensional pupil apodizations for which no specific geometric constraints are put on the pupil plane apodization, apart from the shape of the aperture itself. Masks for circular and segmented apertures are displayed, with and without central obstruction and spiders. Examples of optimal masks are shown for Subaru, SPICA and JWST. Several high-contrast regions are considered with different sizes, positions, shapes and contrasts. It is interesting to note that all the masks that result from these optimizations tend to have a binary transmission profile. 
\end{abstract}




\section{Introduction}

As the population of known exoplanets has grown over the last decade, interest has been steadily increasing in methods and missions to observe them directly.  These include both ground and space based systems.  In fact, the 2010 decadal survey for astronomy and astrophysics \cite{decadal} placed a high priority on developing both the technology and mission concepts for a high-contrast planet imager that would be ready for the next decade.   The goal is to be able to image Earthlike planets in the habitable zones of their parent stars with a telescope from $4$ to $8$ m in diameter.  Many approaches to achieving the high contrast needed have been proposed, including many types of coronagraphs, pupil interferometers, and external occulters.
In this paper, we present new shaped-pupil designs that increase the system throughput and can be used with on-axis and segmented telescopes, introducing the potential for dramatically simplifying the engineering of a future ground or space telescope.

{\em Pupil apodization} creates a high-contrast point-spread function by varying the amplitude transmission in the entrance pupil of the telescope (see, e.g., \cite{jacquinot}).  In fact, Slepian \cite{Slepian1965} found the best possible apodization in the sense that it maximally concentrates the light in the image plane toward the center leaving very high contrast outside the inner working angle.  Unfortunately, manufacturing aperture masks with variable transmission over broad bands with sufficient accuracy has proved formidable.  Instead, we have focused on binary masks that reshape the entrance pupil by either allowing or blocking transmission \cite{Spergel2001}.

Our approach to designing shaped pupils has  been to maximize the transmission of the pupil subject to constraints on the field distribution in the image plane.
Because a linear operator (the Fourier transform) is used to switch from one plane to the other, these optimization problems are linear programs, and can thus be solved very efficiently. However, so far, full optimizations (i.e. without any constraints put on the pupil except for the initial transmission of the aperture) have only been done for one-dimensional geometries. The results of these optimizations include barcode masks, checkerboard masks and concentric ring masks \cite{Vanderbei2003, Kasdin2003, Vanderbei2004}. In the last two cases, a square geometry and a circular symmetric geometry, respectively, were assumed so that these 1D problems resulted in 2D masks.

An important observation is that each of these
1D optimizations tended toward binary distributions of amplitude in the pupil plane; their binary aspect is not a constraint but rather is a result of the optimization. The outer working angle (OWA) of these masks govern (indirectly) the number of the blocking and transmitting regions, and in the case of a  large OWA the width of these regions becomes smaller than the width associated with one element of resolution. In this case the apodization profile is no longer binary but rather smooth.

In this paper, we describe a new optimization approach that results directly in 2D shaped pupils with no simplifying geometric assumptions. The idea, first described in \cite{Vanderbei2011}, exploits the main idea behind the fast Fourier transform (see also \cite{Soummer2007}) thereby allowing us to design masks for arbitrary pupil geometries, including those with central obstructions, spiders, or segmented telescopes.  It also allows us to tailor the region of high contrast in the image plane to any desired shape or size.  Thus, for instance, we can design for dark holes that match those created by the accompanying wavefront control system.  This ability often results in systems with higher transmission, smaller inner working angles (IWA) or higher contrast.

The paper is organized as follows.  In Section \ref{sec1}, we describe the formalism for the optimizations. In Section \ref{sec2}, we present the result of the optimization for a circular unobstructed aperture. We use this example to show how the sampling of the pupil plane affects the quality of the optimization both in the pupil plane and the image plane. In Section \ref{sec3}, we display various optimized circular pupil masks with a central obscuration and spiders. For example, we consider the pupil geometries of the 8m Subaru Telescope
and the 3m telescope for the planned Japanese Space Infrared Telescope for Cosmology and Astrophysics (SPICA).
Section \ref{sec4} shows that we can also optimize for a segmented mirror; we use the pupil of the James Webb Space Telescope (JWST) 
as an example for such masks. Section \ref{sec5} discusses the possibilities offered by these optimizations, as well as their limitations.

\section{Formalism of the two dimensional pupil optimization}
\label{sec1}

What makes these optimizations possible is a fast and accurate computation of the Fourier Transform (FT) of the pupil plane apodization. Indeed, the amplitude distribution in the image plane is proportional to the Fourier transform of the electric field in the pupil plane, which is the arriving electric field (to be of constant, say unit, amplitude) multiplied by the apodization function $A(x,y)$:
\begin{equation}
E(u,v)=\frac{e^{2 i \pi \frac{F}{\lambda}}}{i \lambda F} \int_{-D/2}^{D/2} \int_{-D/2}^{D/2} A(x,y) e^{-2 i \pi \frac{x u + y v}{\lambda F}} dx dy.  \label{1}
\end{equation}
Here, the aperture has diameter $D$ along the $x$ and $y$ axes (the aperture is therefore assumed to be contained within a $D \times D$ square), the focal length of the instrument is $F$, and $\lambda$ denotes the light's wavelength. 

An important first step in many optical design problems is to replace quantities representing physical lengths with quantities that are unitless scalars times some natural physical scale for the problem.  For example, in the pupil plane we replace $x$ and $y$ with new variables, for which we use the same notation, that measure length in units of $D$ (formally, $x \rightarrow xD$ and $y \rightarrow yD$).   That is, $x=1$ means one diameter.   Similarly, in the image plane we replace the physical lengths $u$ and $v$ with values that represent radians on the sky (formally, $u \rightarrow u \lambda F/D$ and $v \rightarrow v \lambda F/D$).   With this substitution, an image plane variable $u$ represents a physical position of $u \lambda F/D$ on the detector.  It also represents an angular position of $u \lambda/D$ radians on the sky.  Throughout the rest of this paper we view image plane variables as measures of $\lambda/D$ radians on the sky.  Making these substitutions, (\ref{1}) becomes
\begin{equation}
E(u,v)=\frac{D^2 e^{2 i \pi \frac{F}{\lambda}}}{i \lambda F} \int_{-1/2}^{1/2} \int_{-1/2}^{1/2} A(x,y) e^{-2 i \pi (x u + y v)} dx dy.  \label{2}
\end{equation}
This two-dimensional Fourier transform can be approximated using discrete sets of points in both the pupil and the image planes:
 
\begin{equation}
E(u_{k},v_{l})=\frac{D^2 e^{2 i \pi \frac{F}{\lambda}}}{i \lambda F} \sum_{i=1}^{N_{x}} \sum_{j=1}^{N_{y}} A(x_{i},y_{j}) e^{-2 i \pi (x_{i} u_{k} + y_{j} v_{l})} \Delta x_{i} \Delta y_{j},
\end{equation}

\noindent with $k$ indexed on the set $\{1,\ldots,M_{u}\}$ and $l$ on the set $\{ 1,\ldots,M_{v} \}$. We will assume equal sizes along the $x$ and $y$ axes, as well as along the $u$ and $v$ axes, making $N_{x}=N_{y}=N$ and $M_{u}=M_{v}=M$. Furthermore, we will consider a uniform partition in the $(x,y)$-plane, and thus $\Delta x_{i}=\Delta y_{i}=1/N \, \forall i$.


For the previous one-dimensional problems (where only single Fourier transforms were considered with the result extended to two dimensions by symmetry), the complexity of the problem was $N M$, where $N$ and $M$ are respectively the number of ``pixels'' in the pupil and image plane. A basic 2D FT leads to a complexity of $N^{2} M^{2}$; for a typical desktop computer, the maximum value for $N M$ is roughly 4500, which puts serious limits on how large $N$ and $M$ can be.  For example, if $M$ is chosen to be 30, which is about the minimum one can get away with in the image plane, then $N$ can be at most 150. Consequently, the smallest detail in a pupil mask of 1 cm in diameter would have a width of $\approx 33 \mu m$. This is too low a resolution to produce sufficient contrast, as explained in more detail in Section~\ref{sec2}.

There is, however, a way to increase the resolution with which we sample the pupil plane. As described in \cite{Soummer2007} and \cite{Vanderbei2011}, the complexity can be reduced to $N^{2} M + N M^{2}$ if the 2D Fourier transform is made in two steps, first along one axis,
\begin{equation}
E_{temp}(u_{k},y_{j}) = \sum_{i=1}^{N} A(x_{i},y_{j}) e^{-2 i \pi x_{i} u_{k}} \Delta x,
\end{equation}
and then along the other one, using as inputs the results of the first computation in the second 1D FT,
\begin{equation}
E(u_{k},v_{l})=\frac{D^2 e^{2 i \pi \frac{F}{\lambda}}}{i \lambda F} \sum_{j=1}^{N} e^{-2 i \pi y_{j} v_{l}} E_{temp}(u_{k},y_{j}) \Delta y.
\end{equation}
This two-step method dramatically improves the complexity problem
by reducing the number of computations at the expense of increased memory allocation (the intermediate matrix $E_{temp}$ has to be stored during this process).

Since most telescope pupils have one or two axes of symmetry, we can use this feature to work with arrays 2 or 4 times smaller. Additionally, because the binary apodization is real, the exponential term in the Fourier transform can be replaced with a cosine. When computing the FT this way, N can be as high as 1000, while M can be set to 50, and the smallest detail in the same 1 cm pupil would now have a $5 \mu m$ length. Note that, in this particular case, we gain a factor $1000^{2} \times 50^{2} / (1000^2 \times 50 + 1000 \times 50^2) \approx 48$ in efficiency over the classical, brute-force, computation of the 2D FT.

The maximization of the pupil mask's throughput, subject to contrast constraints, can be formulated as a linear programming problem: maximize E(0,0) subject to:
\begin{equation}
\begin{split}
-10^{-c/2} & \le \frac{E(u_{k},v_{l})}{E(0,0)} \le 10^{-c/2} \\
 0 & < A(x_{i},y_{j}) < 1
\end{split}
\end{equation}
where $10^{-c}$ is the targeted contrast. The electric field in the image, $E(u_{k},v_{l})$, is subject to this constraint only in a specific discovery space (defined in particular by an IWA and an OWA). The apodization, $A(x_{i},y_{j})$, is also defined on a specific domain, the initial aperture shape.

In practice, the optimization problem is expressed in the AMPL language \cite{Fourer1990} and solved with R. Vanderbei's LOQO solver \cite{Vanderbei1999}. Figure \ref{fig_ampl} shows the AMPL model for the mask shown in Figure \ref{Circular}.

\begin{figure}
\begin{center}
\small
\begin{verbatim}
param pi := 4*atan(1);
param rho0 := 3;	# inner working angle
param rho1 := 15;	# outer working angle
param c := 6; 		# contrast

param N := 500; 	# pupil plane discretization parameter
param dx := 1/(2*N);
param dy := dx;
set Xs := setof {j in 0.5..N-0.5 by 1} j/(2*N);
set Ys := setof {j in 0.5..N-0.5 by 1} j/(2*N);
set Pupil := setof {x in Xs, y in Ys: x^2+y^2 < 0.25} (x,y);

var A {x in Xs, y in Ys: x^2 + y^2 < 0.25} >= 0, <= 1;

param M := 50;		# image plane discretization parameter
set Us := setof {j in 0..M} j*rho1/M;
set Vs := setof {j in 0..M} j*rho1/M;
set DarkHole := setof {u in Us, v in Vs: 
                         u^2+v^2>=rho0^2 
                      && u^2+v^2<=rho1^2  } (u,v);

var C {u in Us, y in Ys};     # this is E_temp
var E {u in Us, v in Vs};

var area = sum {(x,y) in Pupil} A[x,y]*dx*dy;

maximize throughput: area;

subject to C_def {u in Us, y in Ys}:
    C[u,y] = 2*sum {x in Xs: (x,y) in Pupil} 
                       A[x,y]*cos(2*pi*x*u)*dx;

subject to E_def {u in Us, v in Vs}:
    E[u,v] = 2*sum {y in Ys} 
                       C[u,y]*cos(2*pi*y*v)*dy;

subject to sidelobe_pos {(u,v) in DarkHole}:  E[u,v] <= 10^(-c/2)*E[0,0];
subject to sidelobe_neg {(u,v) in DarkHole}: -10^(-c/2)*E[0,0] <= E[u,v];

solve;

printf {(x,y) in Pupil}: "%10f %10f %10f \n", x, y, A[x,y] > "A.out";
\end{verbatim}
\normalsize
\end{center}
\caption{AMPL model for the mask shown in Figure \ref{Circular} with $N = 1000$. }
\label{fig_ampl}
\end{figure}

\section{Circular unobstructed aperture}
\label{sec2}

To begin, we consider a circular unobstructed aperture.  Almost  all coronagraphs being considered for a space mission require unobstructed apertures to achieve high contrast, and therefore necessitate complicated off-axis telescope designs.   Starting with such a geometry allows us to compare the new approach to previously-designed shaped pupils.  In later sections, we explore designs for more interesting pupil geometries.

\subsection{Case of a circular symmetric high contrast region}

In Vanderbei et al.\cite{Vanderbei2003} we found optimal shaped pupils for open circular  apertures consisting of concentric rings by solving for the ring width and spacing along a radius.  This allowed the problem to be solved in only one dimension.  It is, however, interesting to compare the result of the 1D optimization to the new 2D approach. In the latter case, the optimization produces a ``wiggly'' version of a concentric ring mask (the wiggles result from the rather coarse discretization of the image plane). 

\begin{figure}
\begin{center}
\begin{tabular}{cc}
\includegraphics[width=5cm]{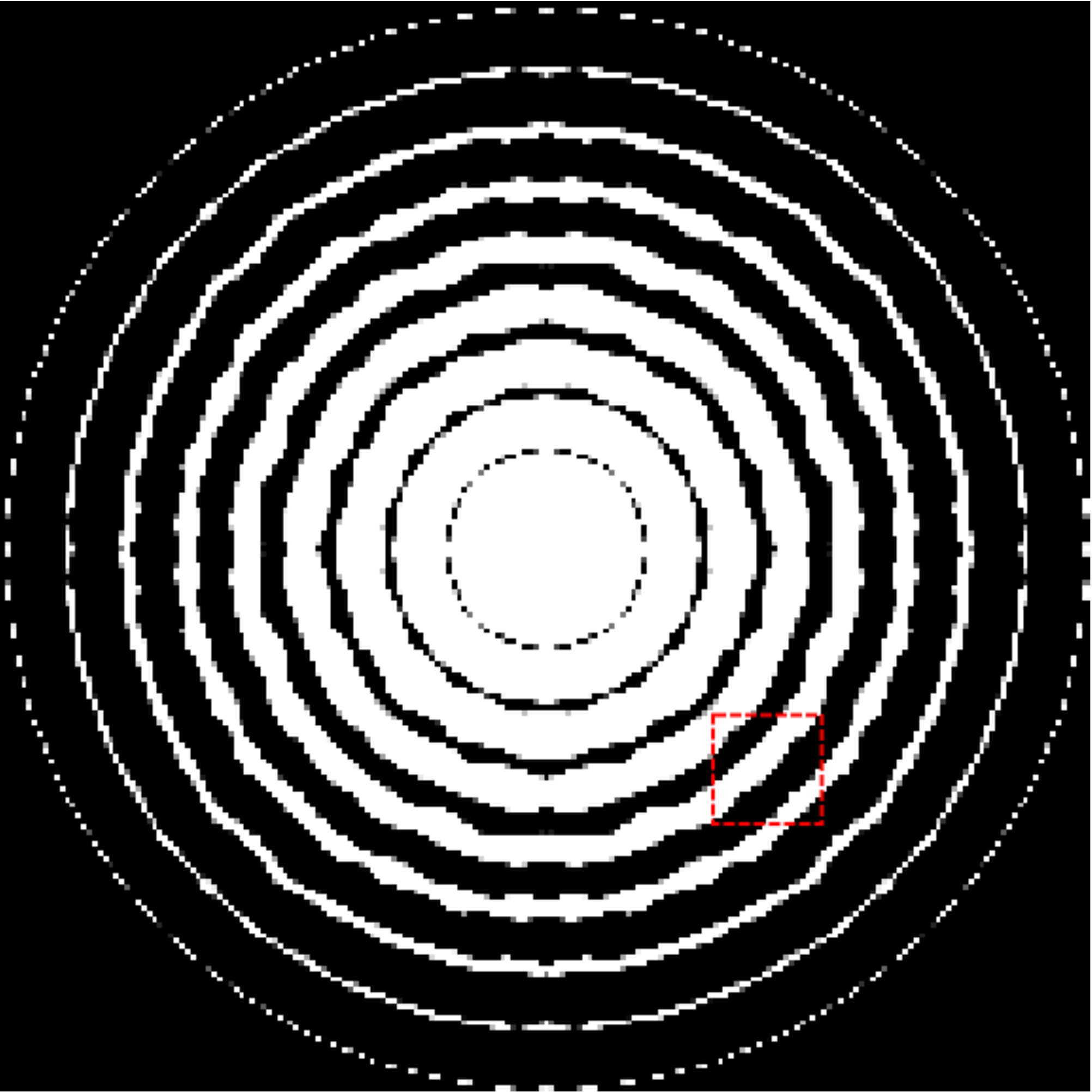} & \includegraphics[width=5cm]{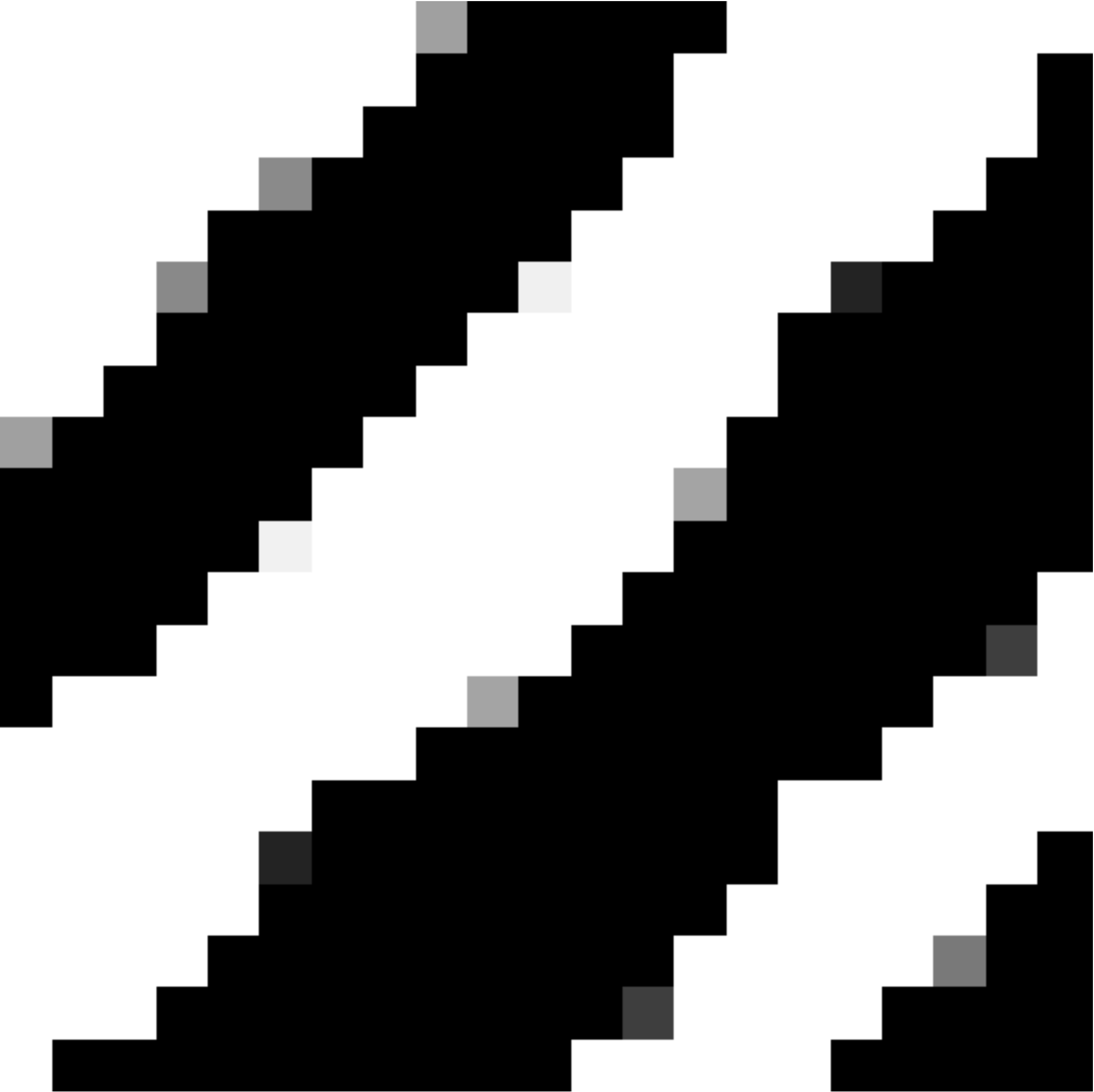}\\
\includegraphics[width=5cm]{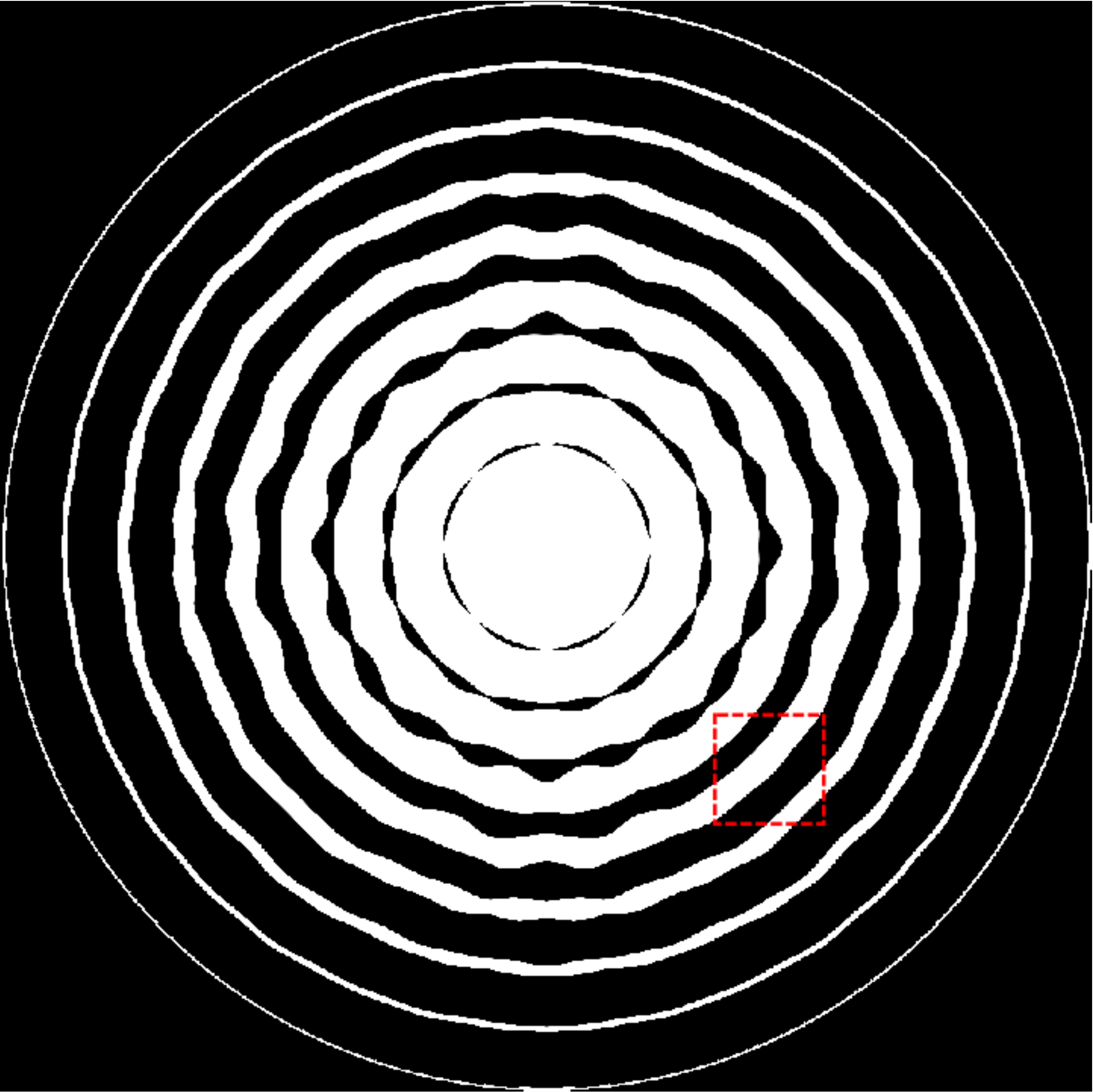} & \includegraphics[width=5cm]{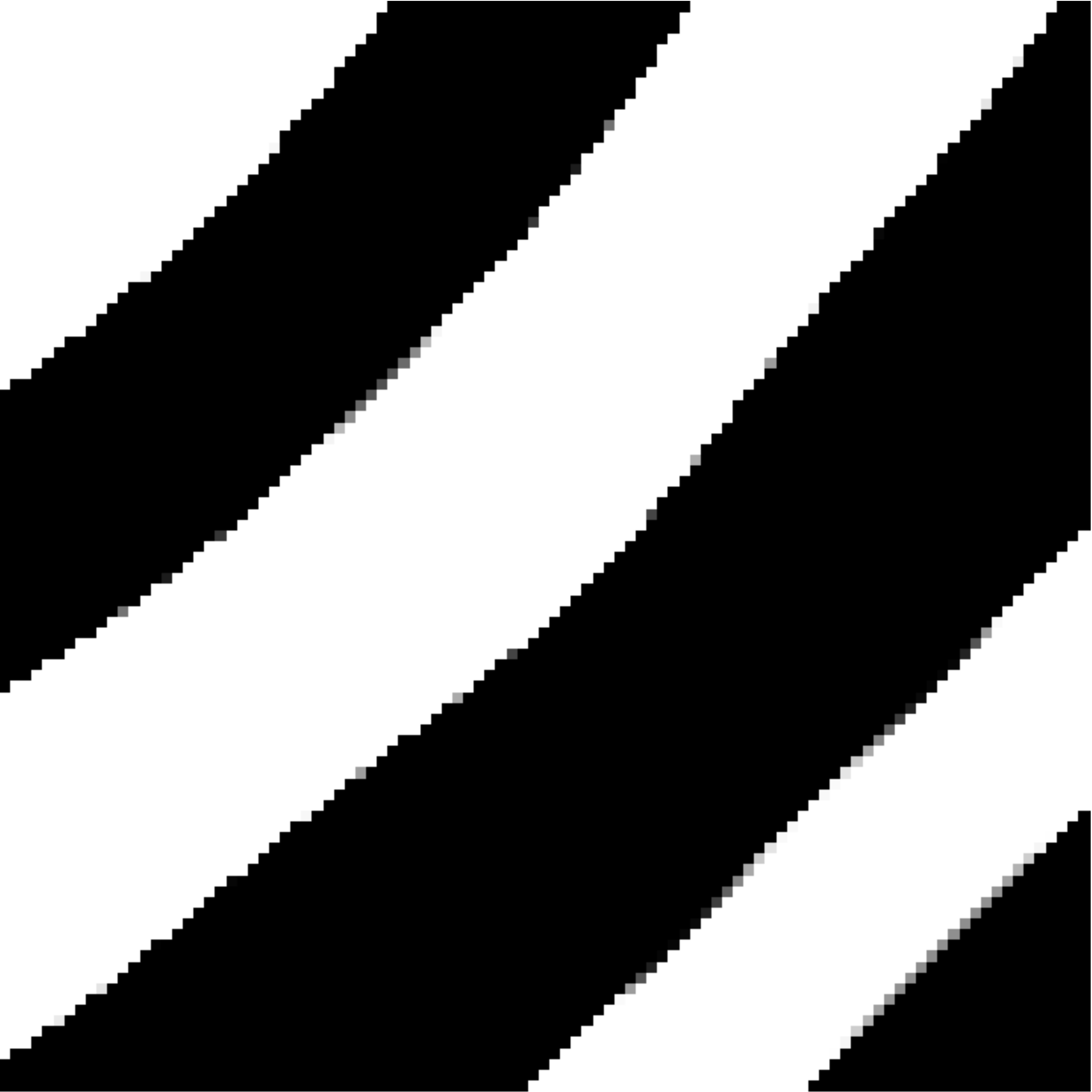}
\end{tabular}
\end{center}
\caption{Pupil mask with a transmission of 36\%. Top: N=200. Bottom: N=1000 (in both cases M=100). Left: optimal pupil mask for an IWA of 3 $\lambda/D$ and an OWA of 15 $\lambda/D$. Right: magnified region contained in the red dashed rectangle displayed on the left. }
\label{Circular}
\end{figure}

\begin{figure}
\begin{center}
\includegraphics[width=12cm]{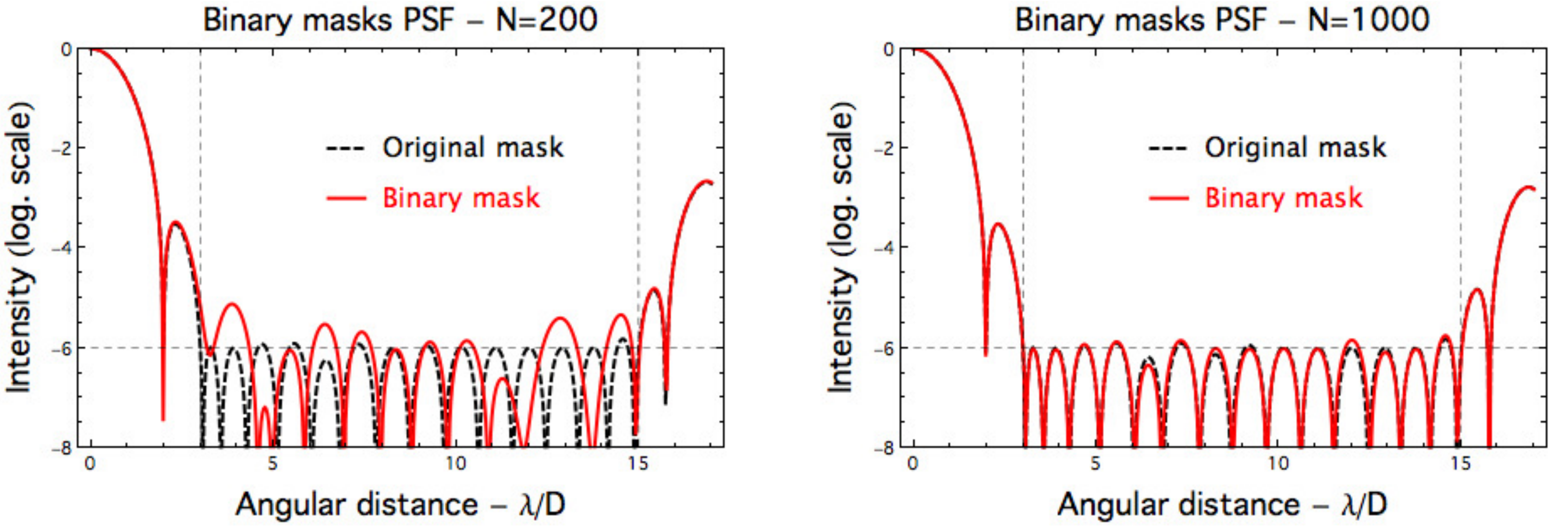}
\end{center}
\caption{Cross section of the PSFs of the masks displayed on Figure \ref{Circular} (left: N=200, right: N=1000) before and after the transmission values have been rounded. The PSF of the optimal mask is represented by a dashed line while a red solid line is used for the binary version of the same mask. In both cases the intensity is represented on a logarithmic scale.}
\label{PSFCircular}
\end{figure}

Figure \ref{Circular} displays the results of the optimization with N=200 and with N=1000 (and M=100 in both cases). In fact the optimization uses arrays 4 times smaller, taking advantage of symmetry, but the numbers of points we indicate afterwards are always the total number of points. In both cases the contrast is $10^{-6}$ inside a ring with an IWA of 3 $\lambda/D$ and an OWA of 15 $\lambda/D$, as can be seen in Figure \ref{PSFCircular}. A comparison of the same region is shown; the size of the arrays has a clear influence on the quality of the optimal transmission. In particular, the value of the transmission tends to converge to either 0 or 1. The number of pixels for which the transmission adopts a value between 0 and 1 decreases greatly when the size of the array is increased; the low resolution pupil has $1.8\%$ of its transmitted intensity between $0.1$ and $0.9$, while less than $0.2\%$ of the high resolution pupil has this property.

Figure \ref{PSFCircular} shows a comparison between the PSFs of the pupils displayed in Figure~\ref{Circular} with and without the transmission values rounded. In the former case, the pupil becomes a true binary mask and some unwanted features brighter than $10^{-6}$ can be seen inside the high contrast ring when the pupil is too poorly sampled (N=200). The same does not occur with the higher resolution (N=1000). All of the PSFs that we show in the rest of the paper are computed for the binary masks obtained by rounding all pixels either to zero or to one.

\begin{figure}
\begin{center}
\begin{tabular}{ccc}
\includegraphics[width=3.5cm]{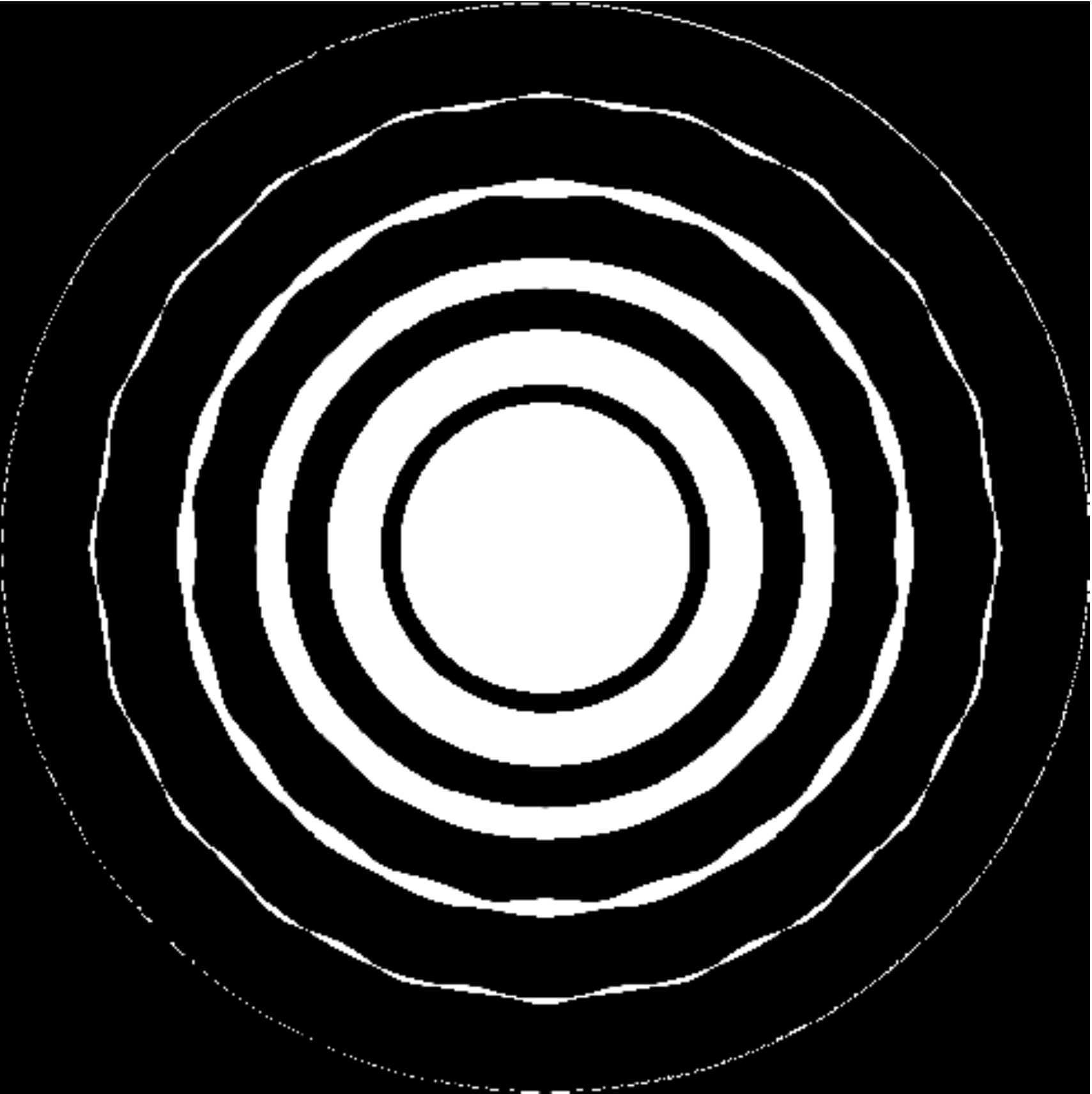} & \includegraphics[width=3.5cm]{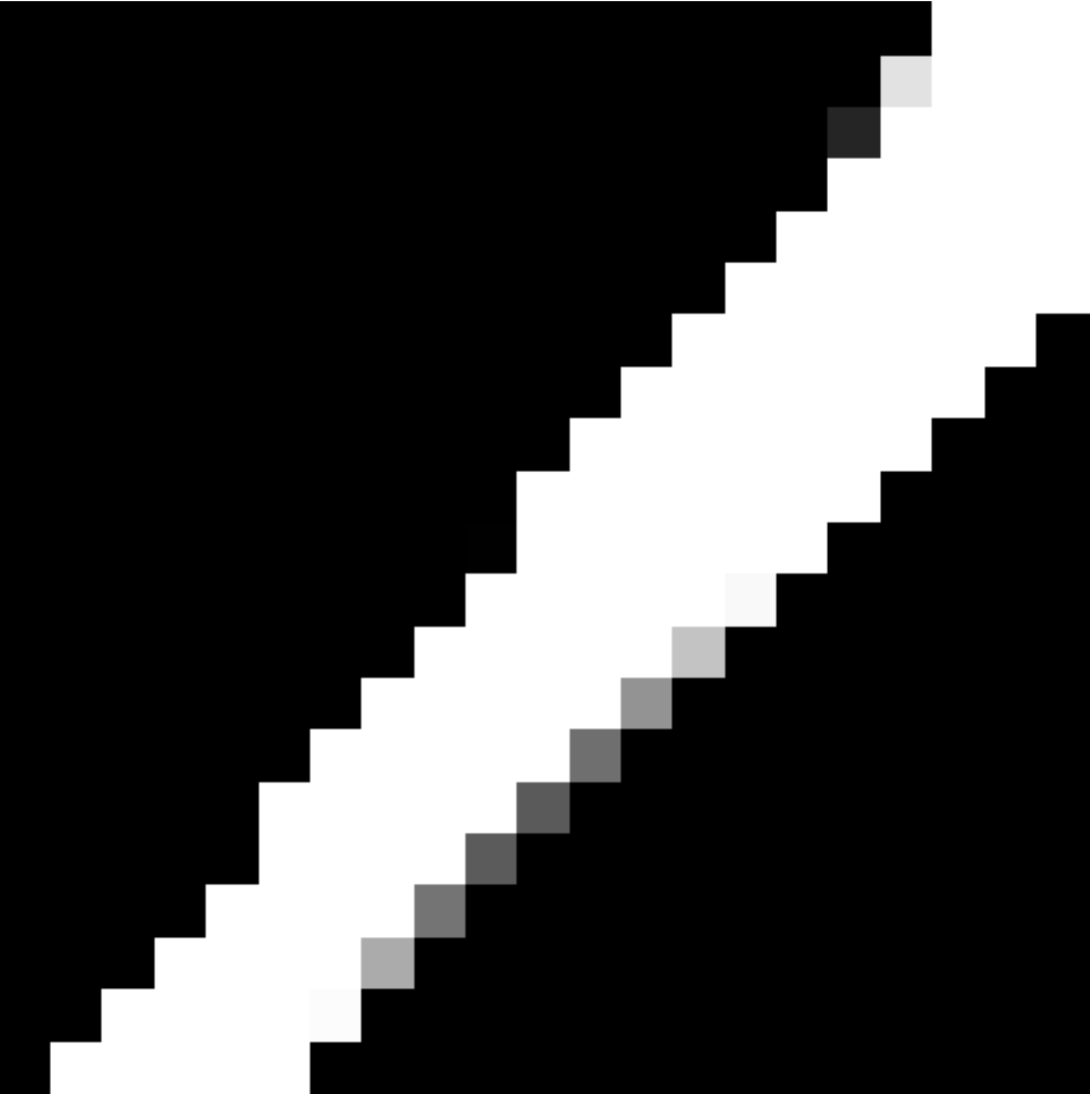} & \includegraphics[width=3.5cm]{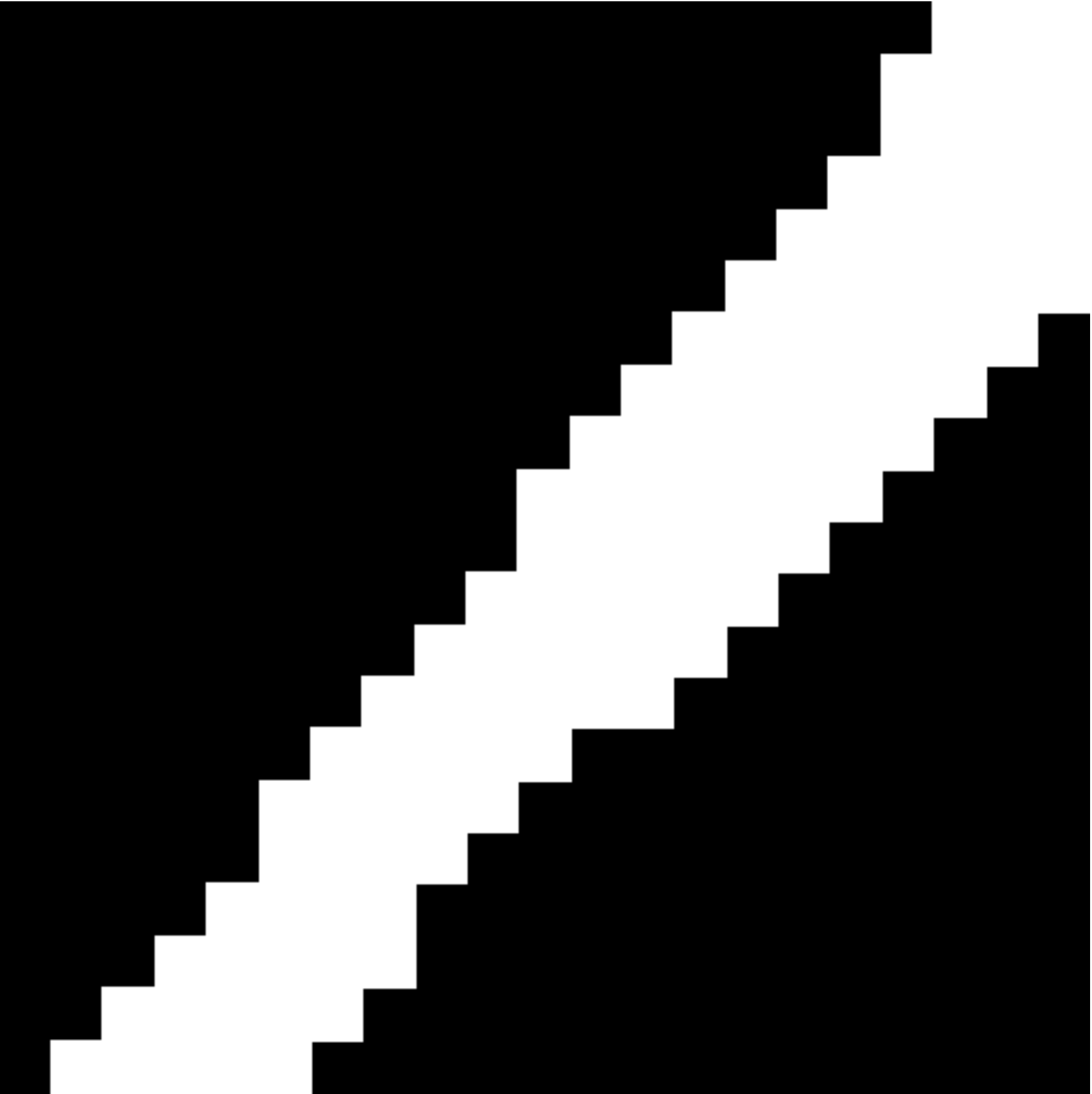} \\
\end{tabular}
\end{center}
\caption{Left: Pupil mask designed for a contrast of $10^{-10}$. Optimization parameters are N=1000, M=100. The throughput is 25\%. Center and right: Detail of the mask before and after rounding the transmission.}
\label{Circular10Pupil}
\end{figure}

\begin{figure}
\begin{center}
\begin{tabular}{ccc}
\includegraphics[width=5cm]{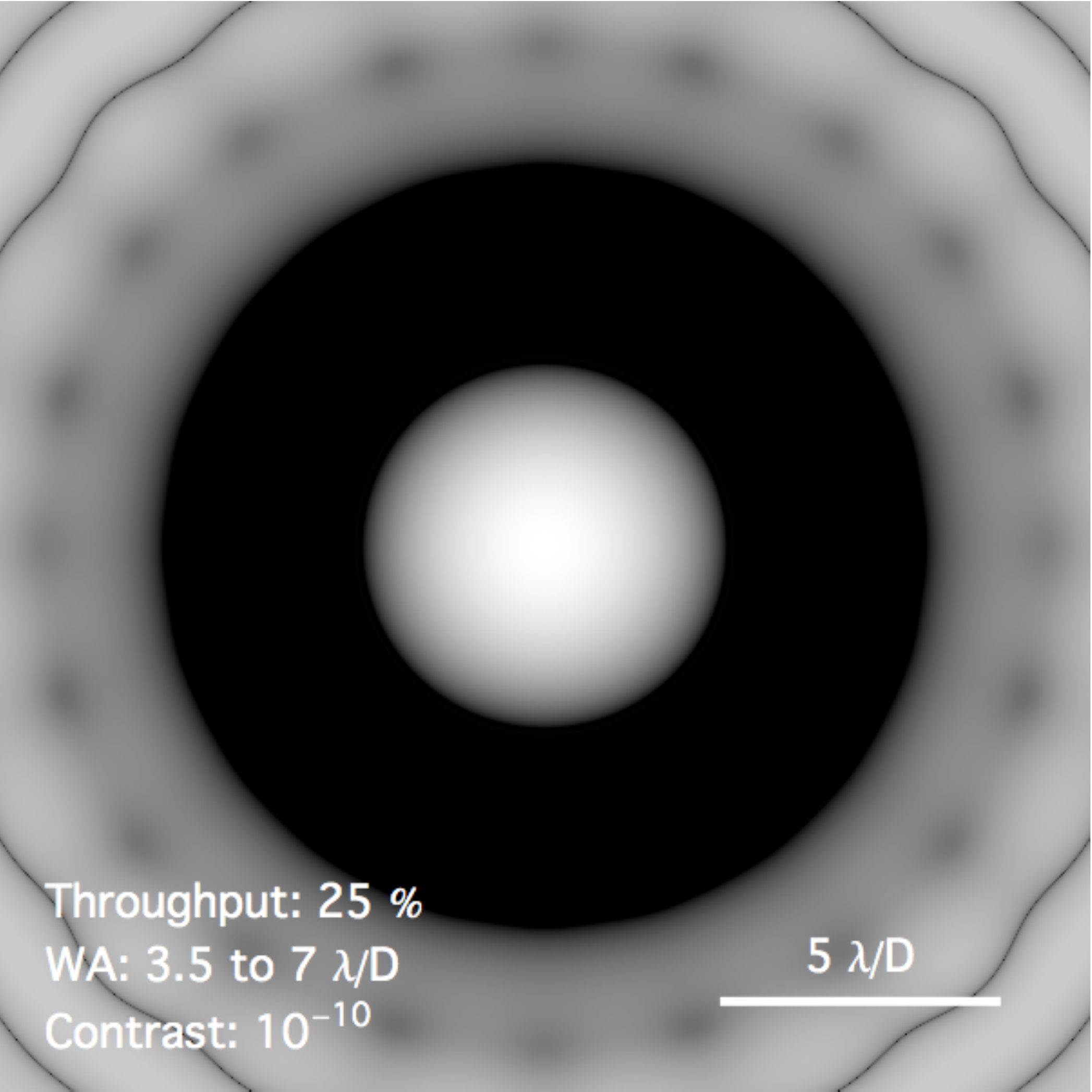} & \includegraphics[width=5cm]{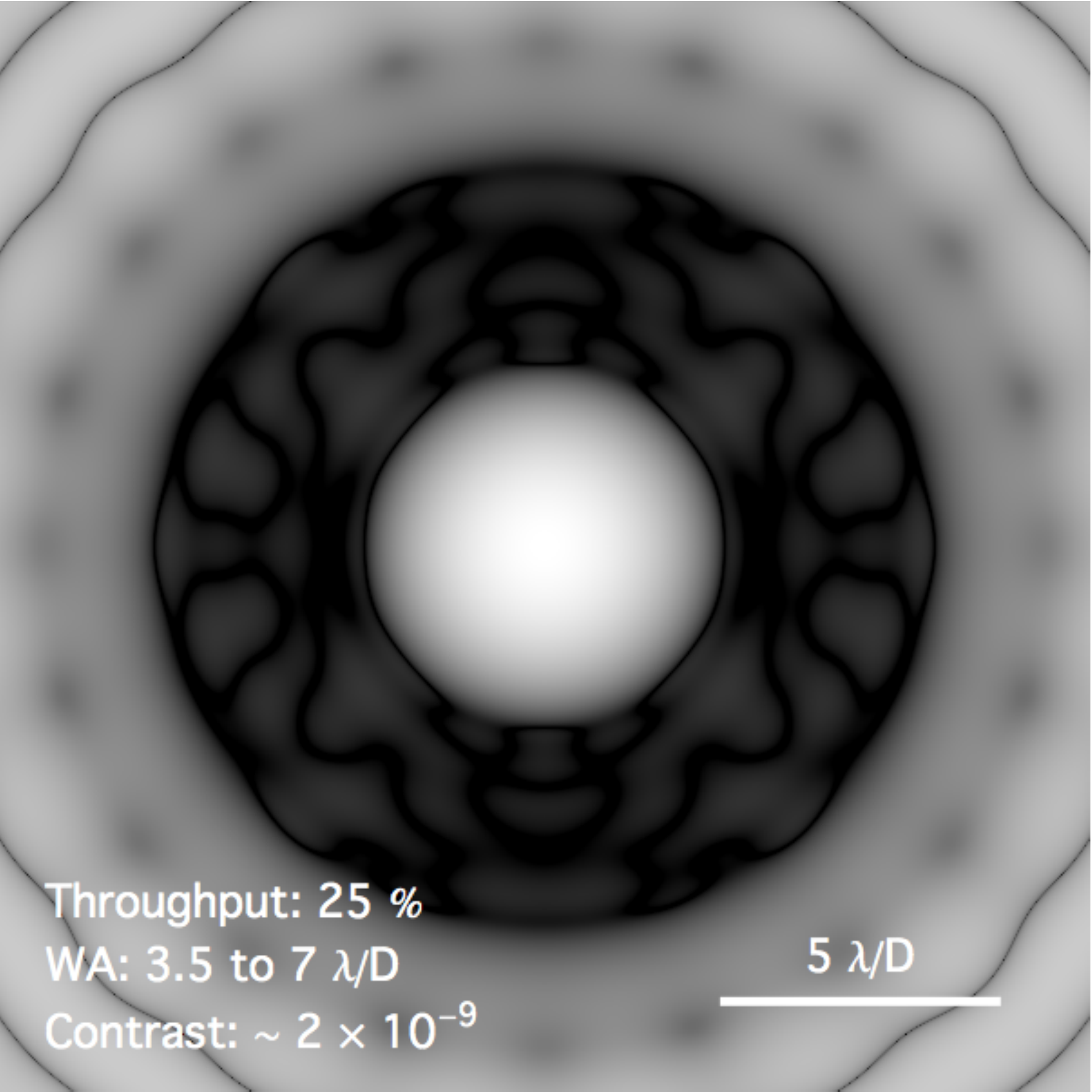} & \includegraphics[width=1.1cm]{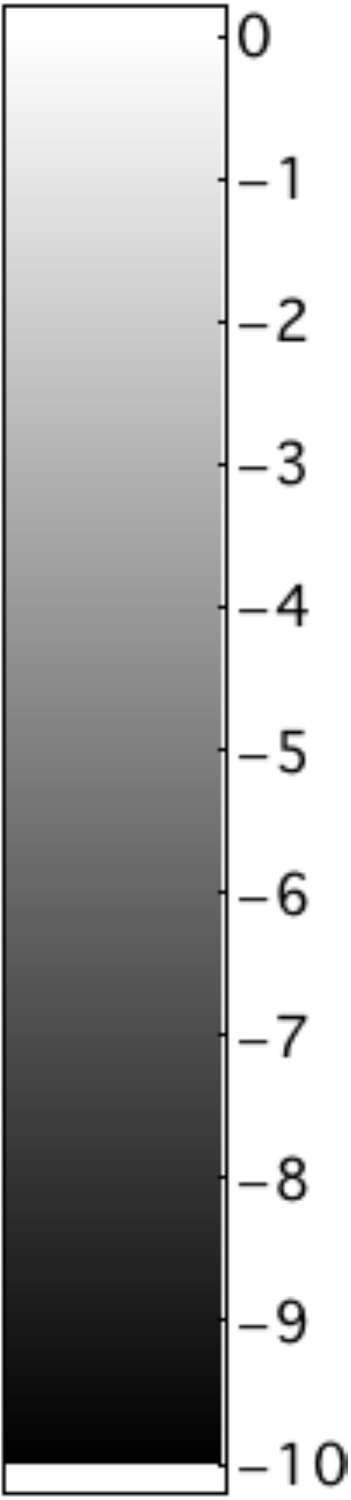}
\end{tabular}
\end{center}
\caption{PSF of the mask displayed on Figure \ref{Circular10Pupil} without (left) and with (right) an artificial rounding of the transmission values of the mask. The IWA is 3.5 $\lambda/D$ and the OWA is 7 $\lambda/D$.}
\label{Circular10PSF}
\end{figure}

There is, however, one exception. We do the comparison between a true concentric ring mask (illustrated in Figure 4 of \cite{Vanderbei2003}) and a similar mask (see Figure \ref{Circular10Pupil}) fully optimized in 2D for the same IWA and OWA (3.5 and 7 $\lambda/D$) and the same contrast ($10^{-10}$). The original concentric ring mask and the new one have similar throughputs (23.5\% instead of 25\%) but the number of rings is not the same in the two cases (5 instead of 6). As seen in Figure \ref{Circular10PSF}, the mask (designed with N=1000, M=100), only provides a mean contrast of $2 \times 10^{-9}$ when its transmission values have been rounded. Less than $0.12\%$ of the pixels have an apodization value between $0.1$ and $0.9$, but the targeted high contrast level makes this small ratio significant enough to change the effective contrast by more than an order of magnitude.  We should, however, point out that the 1D mask was designed using a two-step process.   In the first step, a linear programming problem was solved to find a discretized approximate solution to the problem.   For the second step, the on-off and off-on thresholds from the approximate solution were used as input to a 1D nonconvex nonlinear optimization problem that hones these thresholds to highly precise values.   We have not adapted this second stage to the current 2D design process.  We are currently working on such an enhancement.  

\begin{figure}
\begin{center}
\begin{tabular}{cc}
\includegraphics[width=5cm]{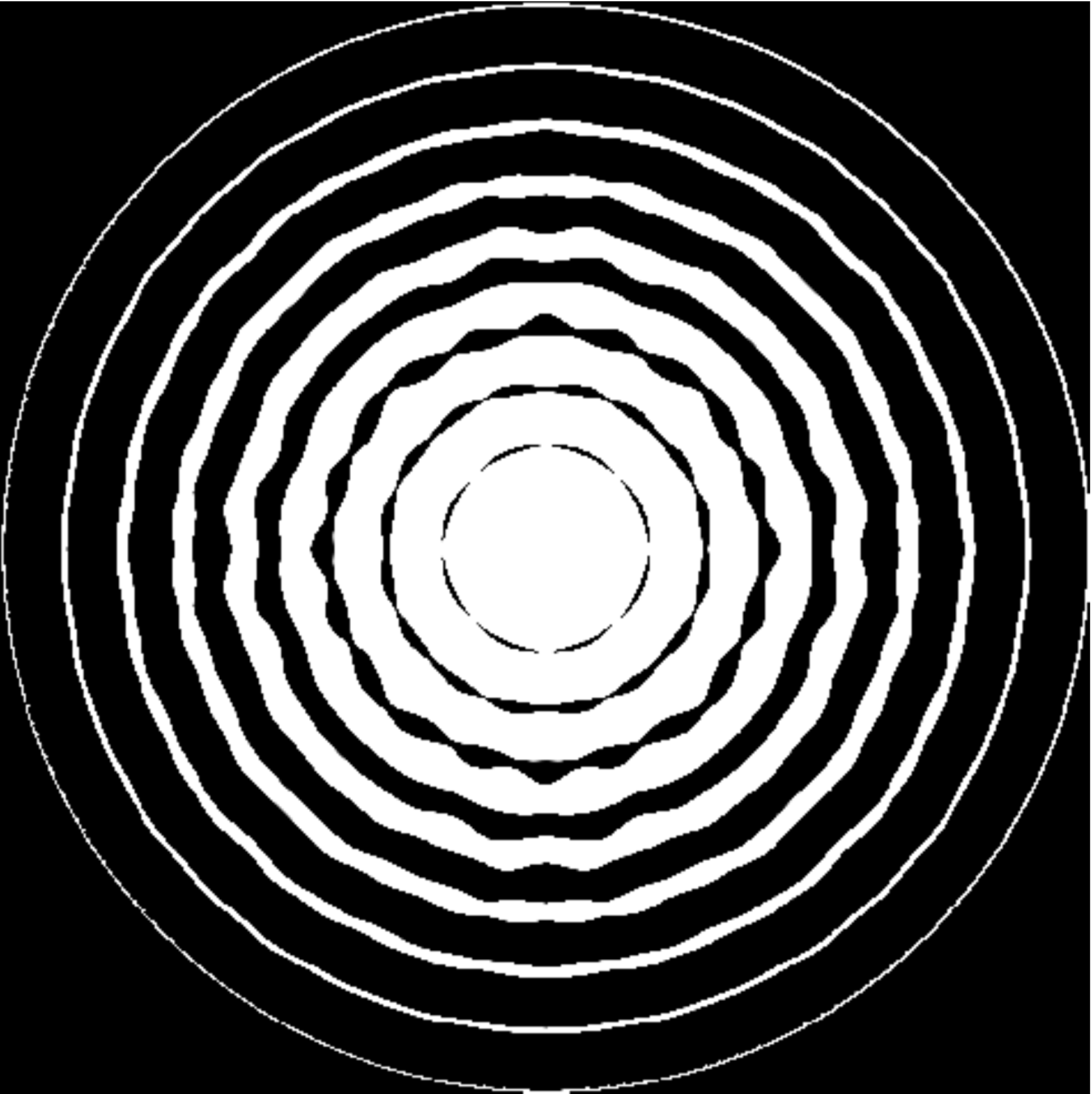} & \includegraphics[width=5cm]{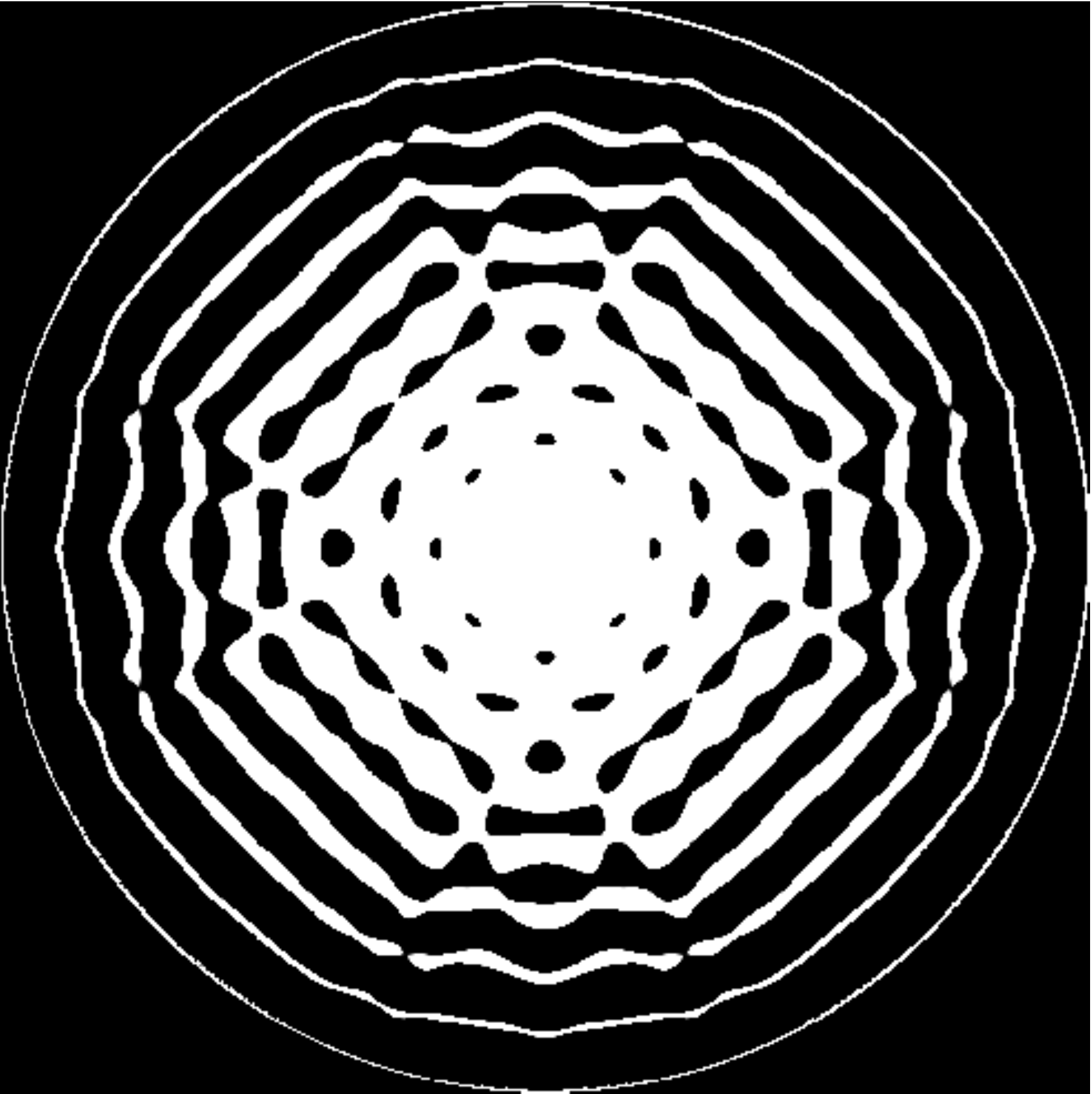}
\end{tabular}
\end{center}
\caption{Two pupil masks optimized with a different sampling precision of the image plane (N=1000 in both cases while M=100 on the left, and 50 on the right). Constraints on the contrast and the IWA and OWA are the same as for the pupils in Figure \ref{Circular}.}
\label{SmallSampling}
\end{figure}

It is also noteworthy that the new pupil masks fail to present the same circular symmetry as in the 1D case, although they share strong similarities. This is explained by the fact that the PSF is computed with a small number of points, $M$=100. Figure \ref{SmallSampling} illustrates how this effect is reinforced if $M$ is set to 50.

The transmission of the mask does not significantly change with different values of M and N, as long as the IWA and OWA remain the same. Therefore, it is possible to compute a first mask with a low resolution (in about 15 minutes for N=200, M=50). If the first estimate of the throughput is deemed high enough, a second optimization can be launched for a much larger array (which will take several hours for N=1000, M=50 on a modern desktop computer).

\subsection{Case of two symmetric dark holes}

Instead of an annular dark hole, one may wish to optimize the pupil mask so as to provide high contrast in more spatially-limited regions of the image plane.  This alternative is particularly relevant when the mask is to be used with a dedicated wavefront correction system that can only create small dark holes with high contrast.  In particular, as described in \cite{Kasdin2011}, masks can be designed with dark holes at a lower contrast and then paired with two DMs to produce the remaining contrast.  This pairing results in an overall system with higher throughput and smaller inner working angle.

\begin{figure}
\begin{center}
\begin{tabular}{cc}
\includegraphics[height=5cm]{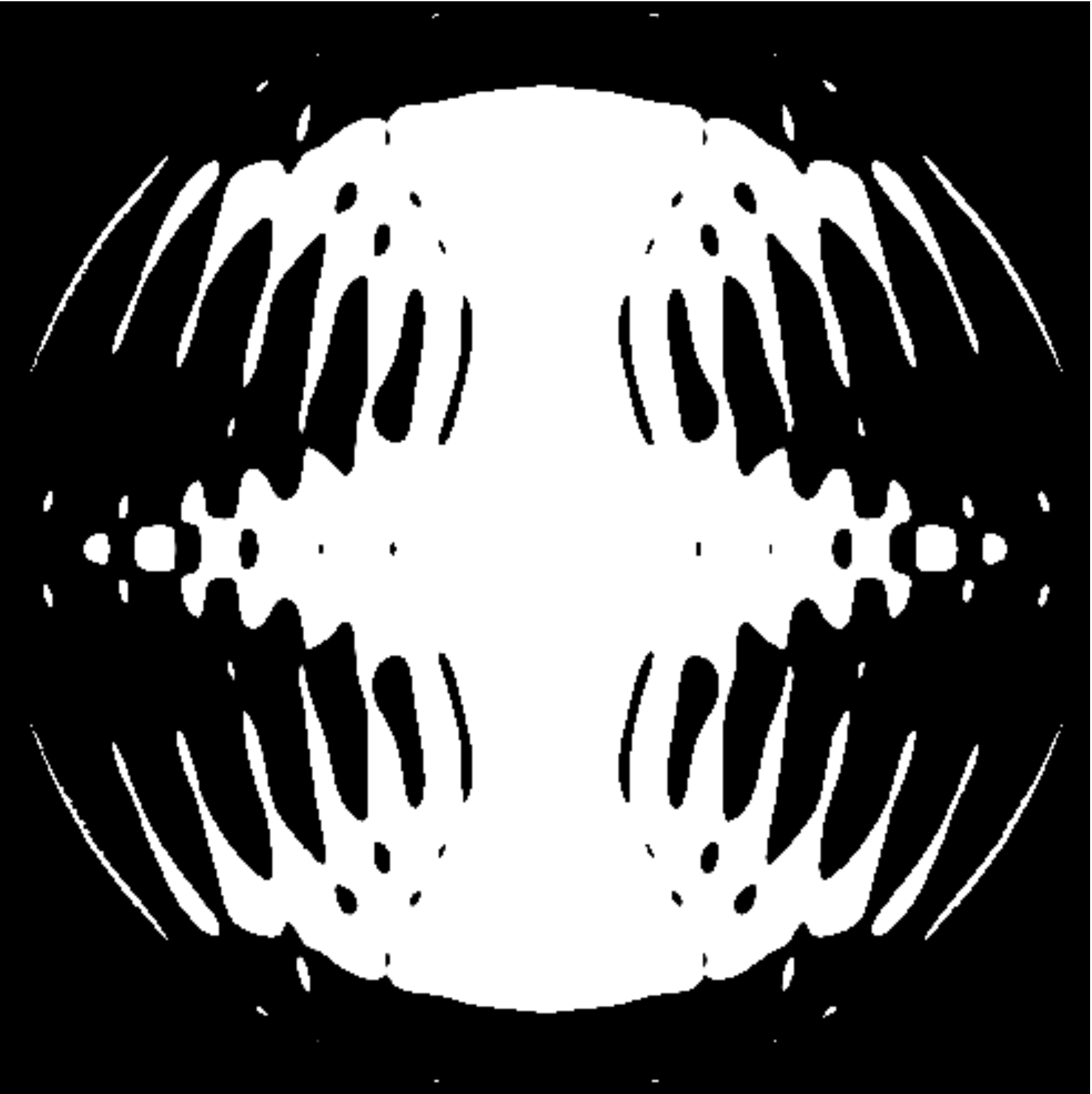} & \includegraphics[height=5cm]{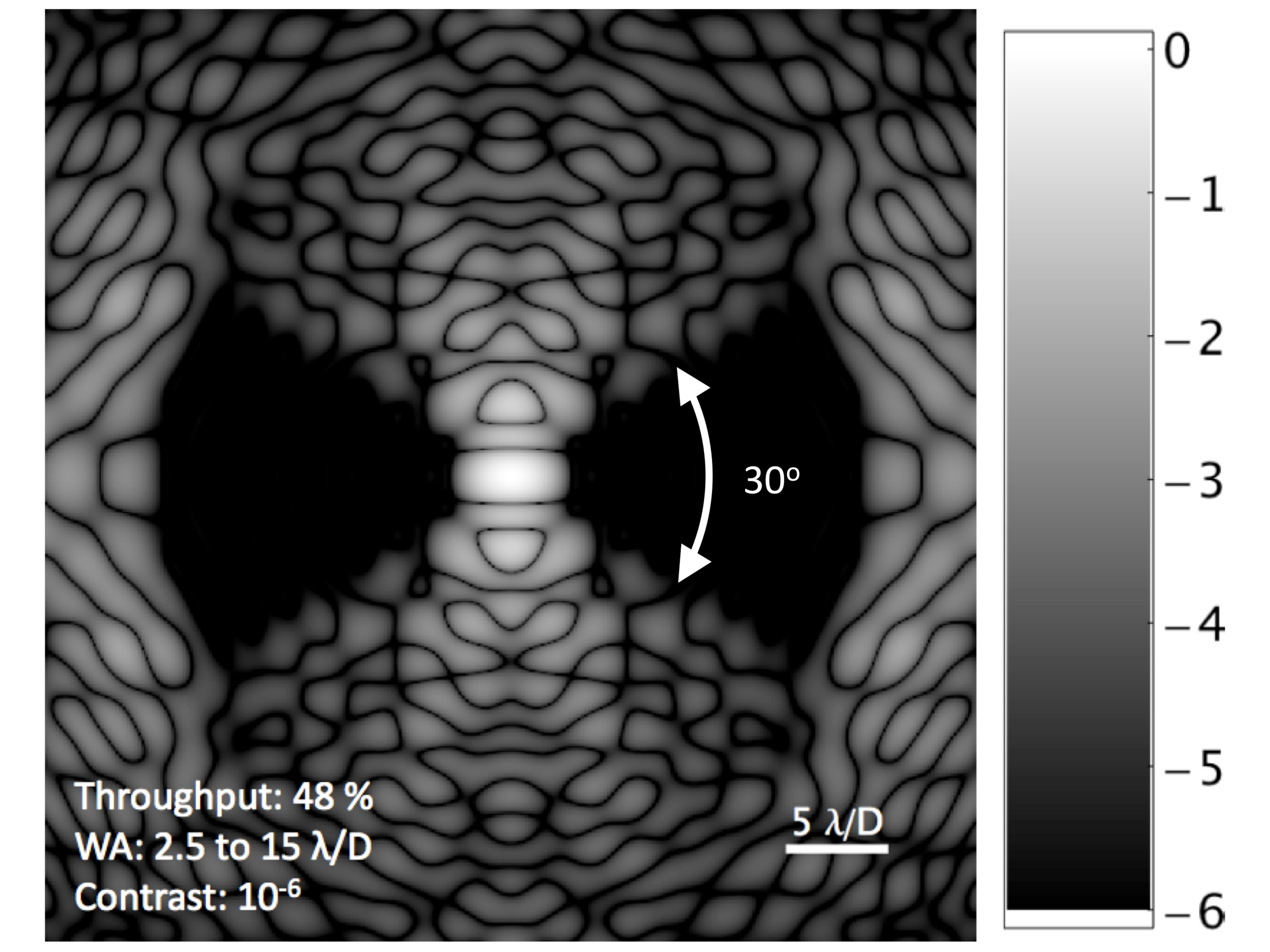}\\
\includegraphics[height=5cm]{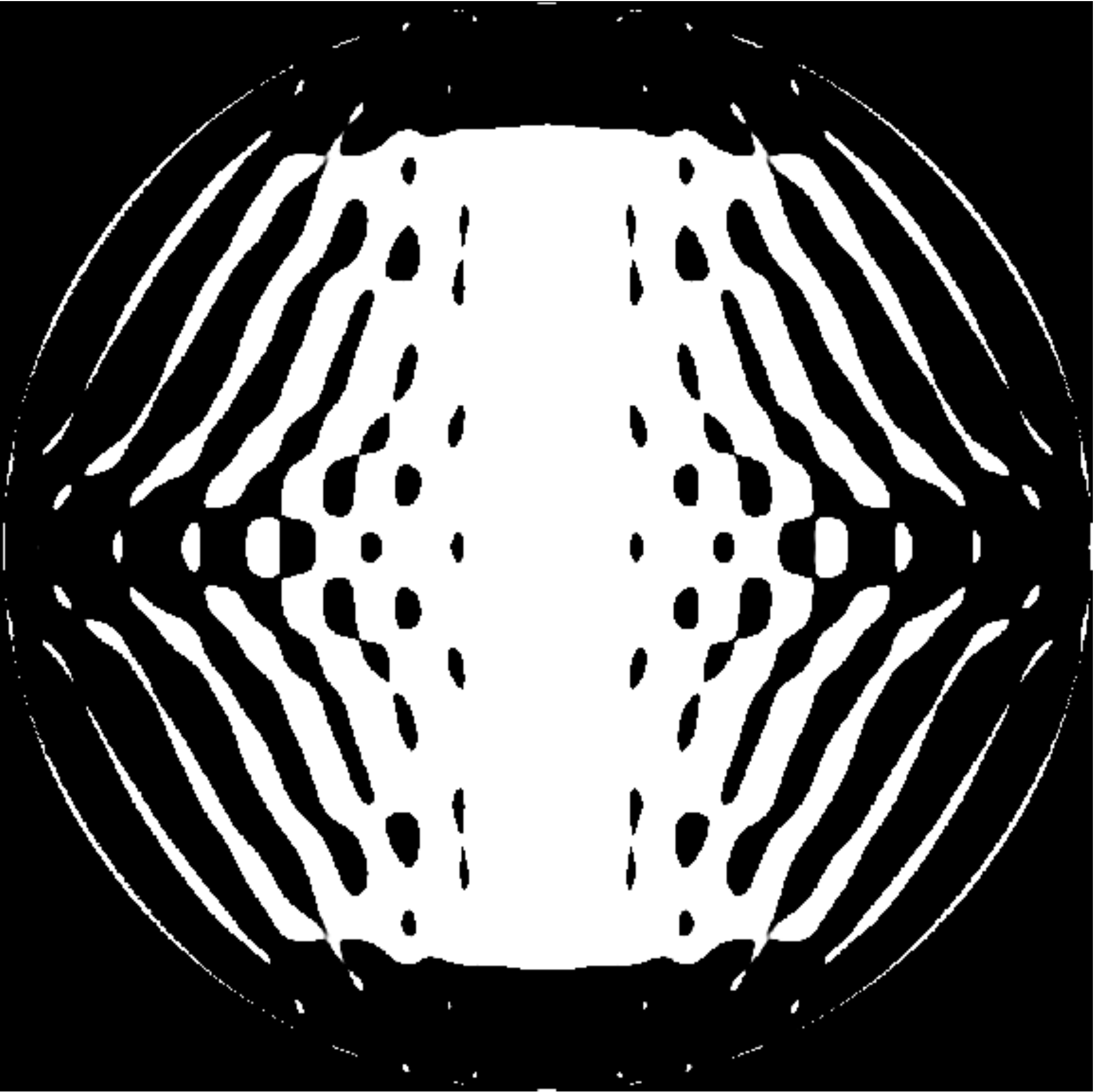} & \includegraphics[height=5cm]{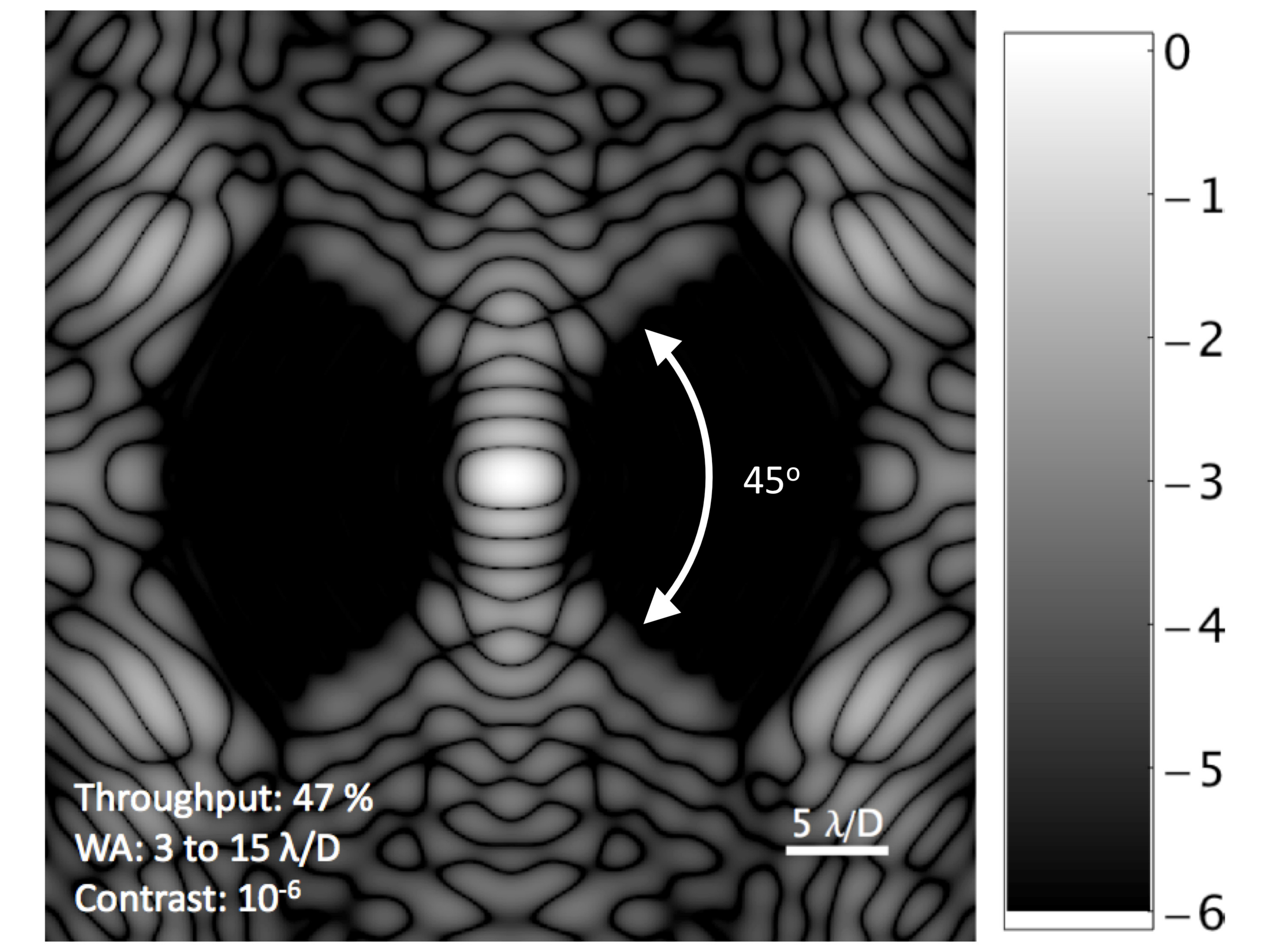}\\
\includegraphics[height=5cm]{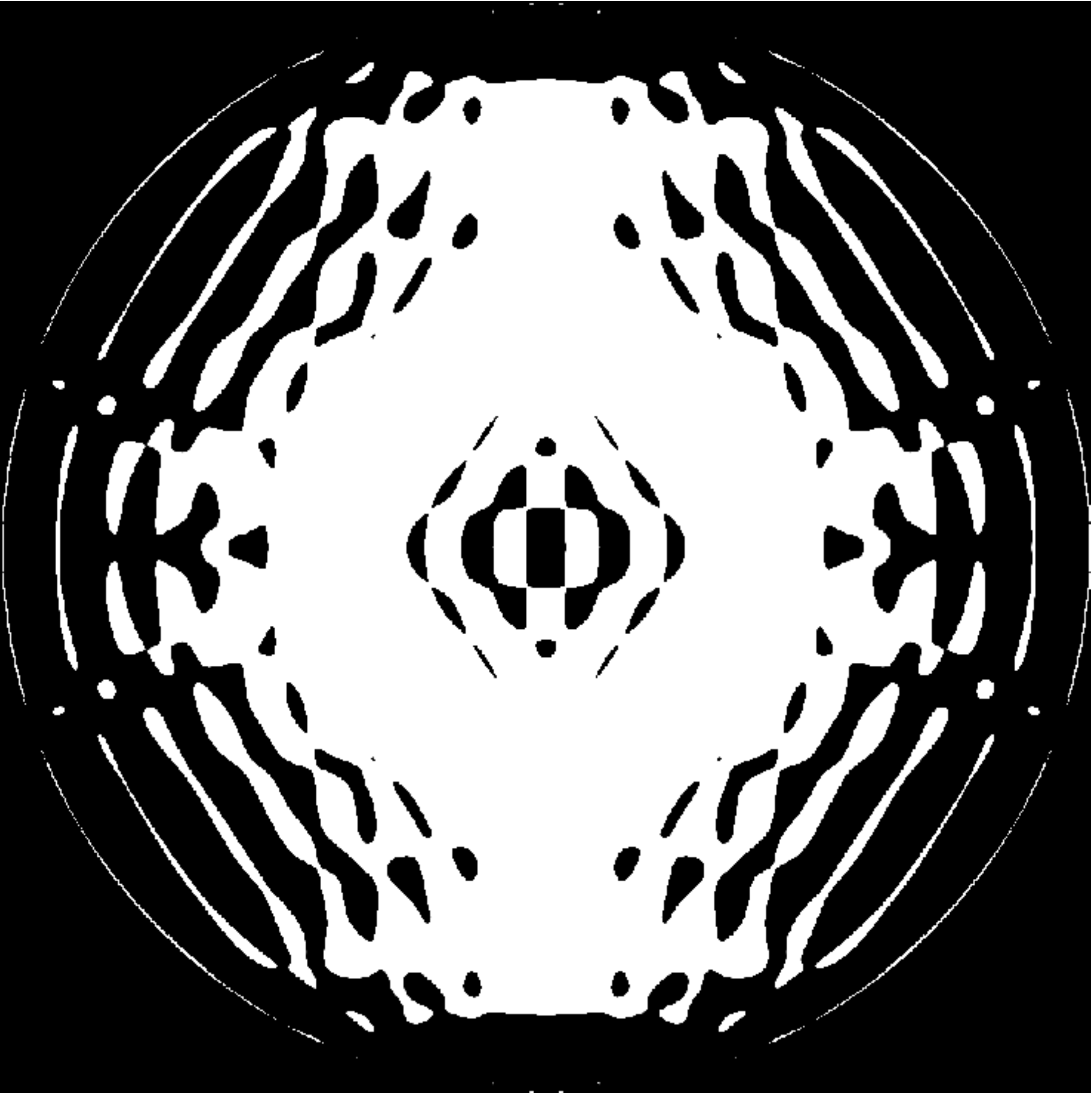} & \includegraphics[height=5cm]{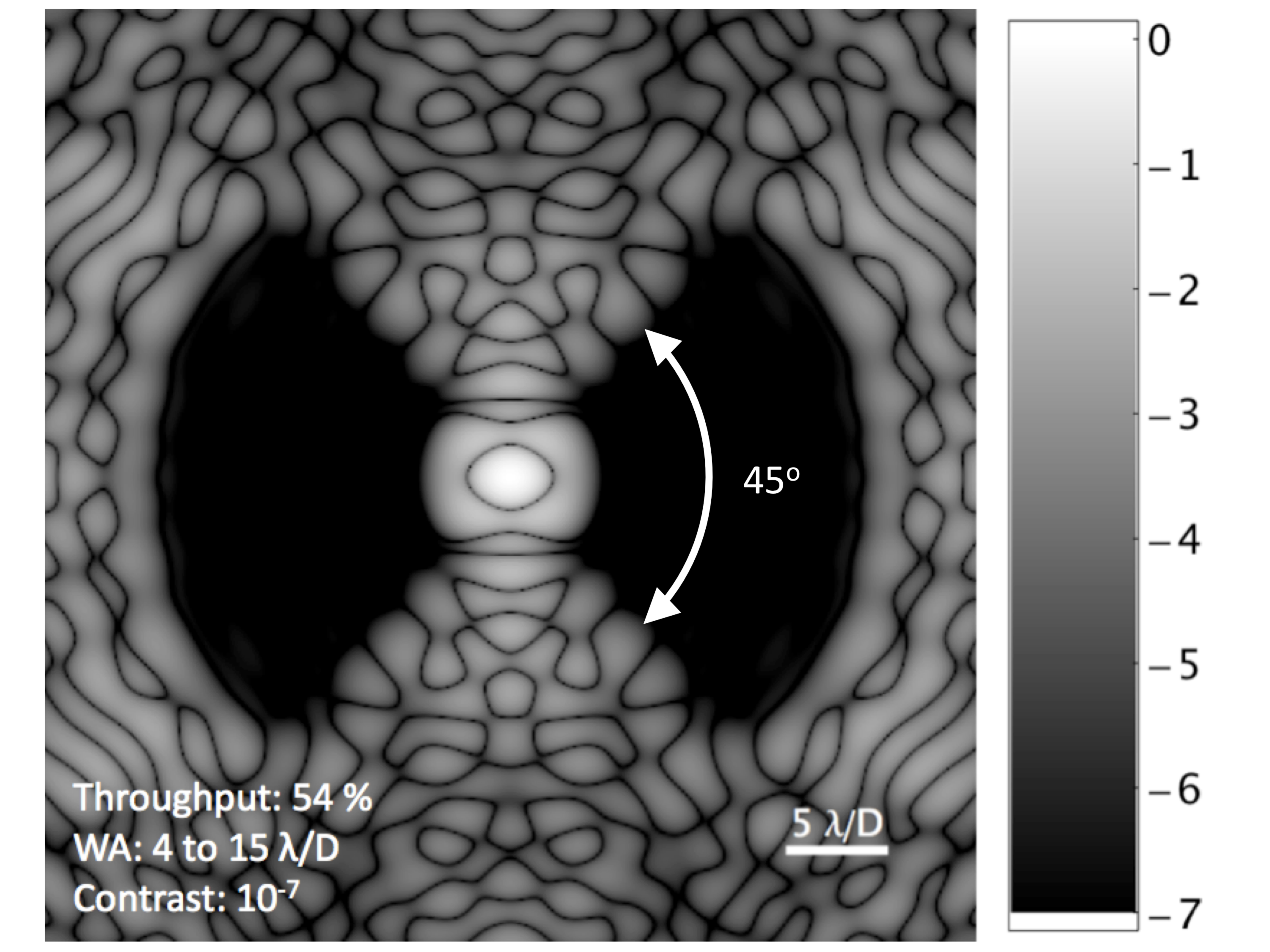}
\end{tabular}
\end{center}
\caption{Pupil masks (left) design to reach $10^{-6}$ (first two rows) and $10^{-7}$ (third row) with a circular unobstructed aperture, and their associated PSFs (right). In both cases the high contrast region is a ring section with an OWA of 15 $\lambda/D$. The IWA are 2.5 $\lambda/D$ (top), 3 $\lambda/D$ (middle) and 4 $\lambda/D$ (bottom). The dark holes have an angular extension of 30 degrees in the first case and 45 degrees in the last two. The PSF are computed for true binary masks, the original transmissions being artificially rounded.}
\label{CircularDH}
\end{figure}

Several tradeoffs exist between the IWA, the throughput and the contrast. Two are illustrated in Figure~\ref{CircularDH}. A comparison between the pupil displayed on the first and second rows indicates that a smaller IWA can be obtained when the size of the dark hole is reduced. The reader will note that, while the throughput remains almost the same, the overall shape of the mask changes dramatically. Furthermore, if a higher contrast is sought, the throughput may be conserved if the IWA is increased, as  is illustrated by the pupils on the first and the third rows. Although not illustrated here, reducing the area of the dark hole may also help increase throughput. 

\section{Circular pupil with a central obstruction and spiders}
\label{sec3}

Since this new two-dimensional approach can create masks for arbitrary pupil geometries, we can, in particular, find masks designed for telescopes with any central obstruction or spider structure.  This allows the coronagraph to maximize the amount of light collected by the telescope. In this section, we present a few masks for two real-world telescope pupils: SPICA and Subaru. These two pupils have two axes of symmetry, allowing us to simplify the computation as explained in section \ref{sec1}.

\subsection{SPICA}

SPICA, the Space Infrared Telescope for Cosmology and Astrophysics, is a planned infrared space telescope consisting of a 3m primary mirror and a secondary mirror of 0.6 m. The central obstruction ratio is then 0.2. The spider's thickness is set to 0.08 m. It should be noted that the design of SPICA is subject to change and the primary and secondary mirrors could respectively have a diameter of 3.2 m and 0.8 m. In the latter case, this would have important consequences on the performances of the coronagraph in terms of throughput, IWA and contrast.

\begin{figure}
\begin{center}
\begin{tabular}{cc}
\includegraphics[height=5cm]{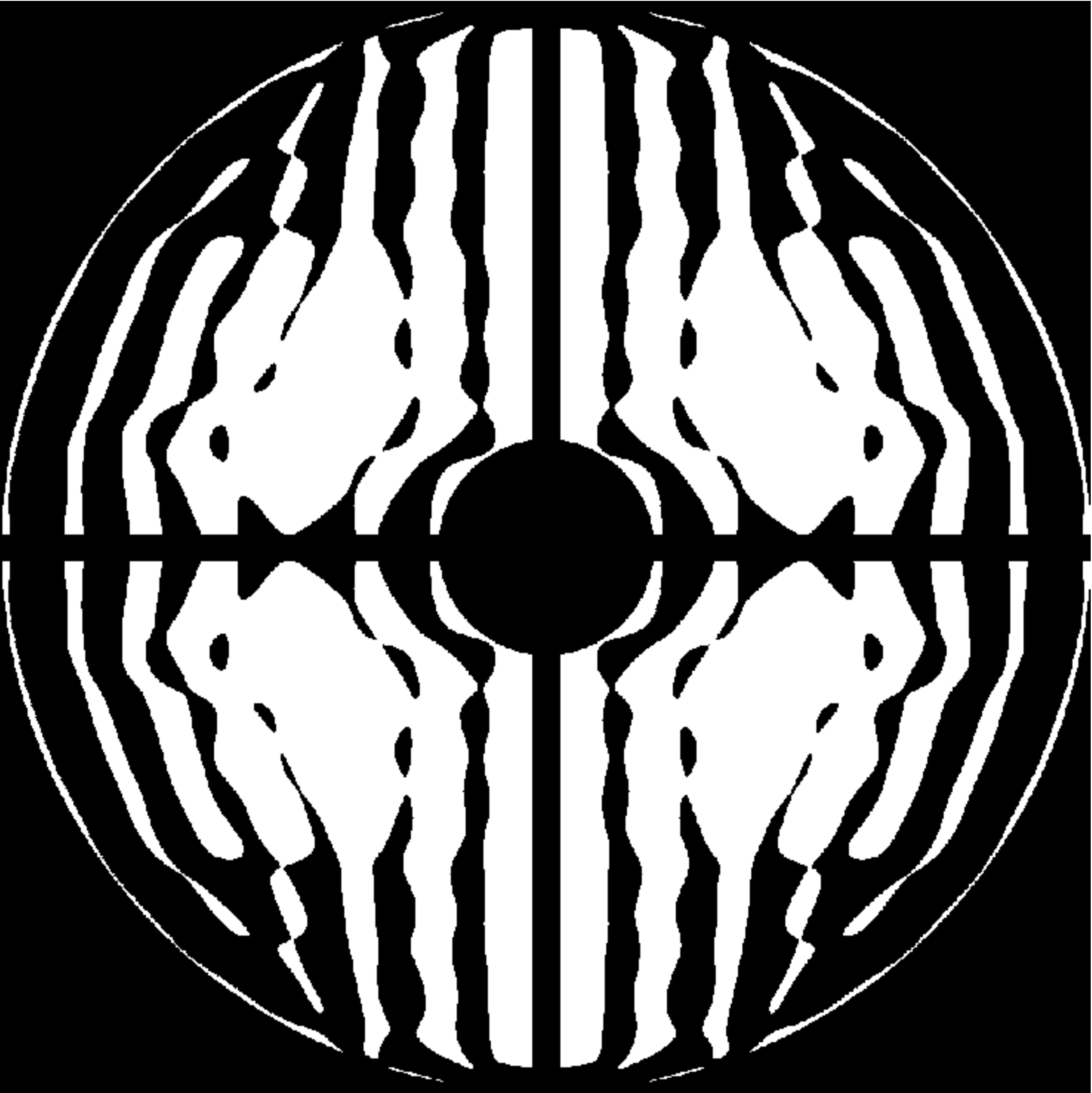} & \includegraphics[height=5cm]{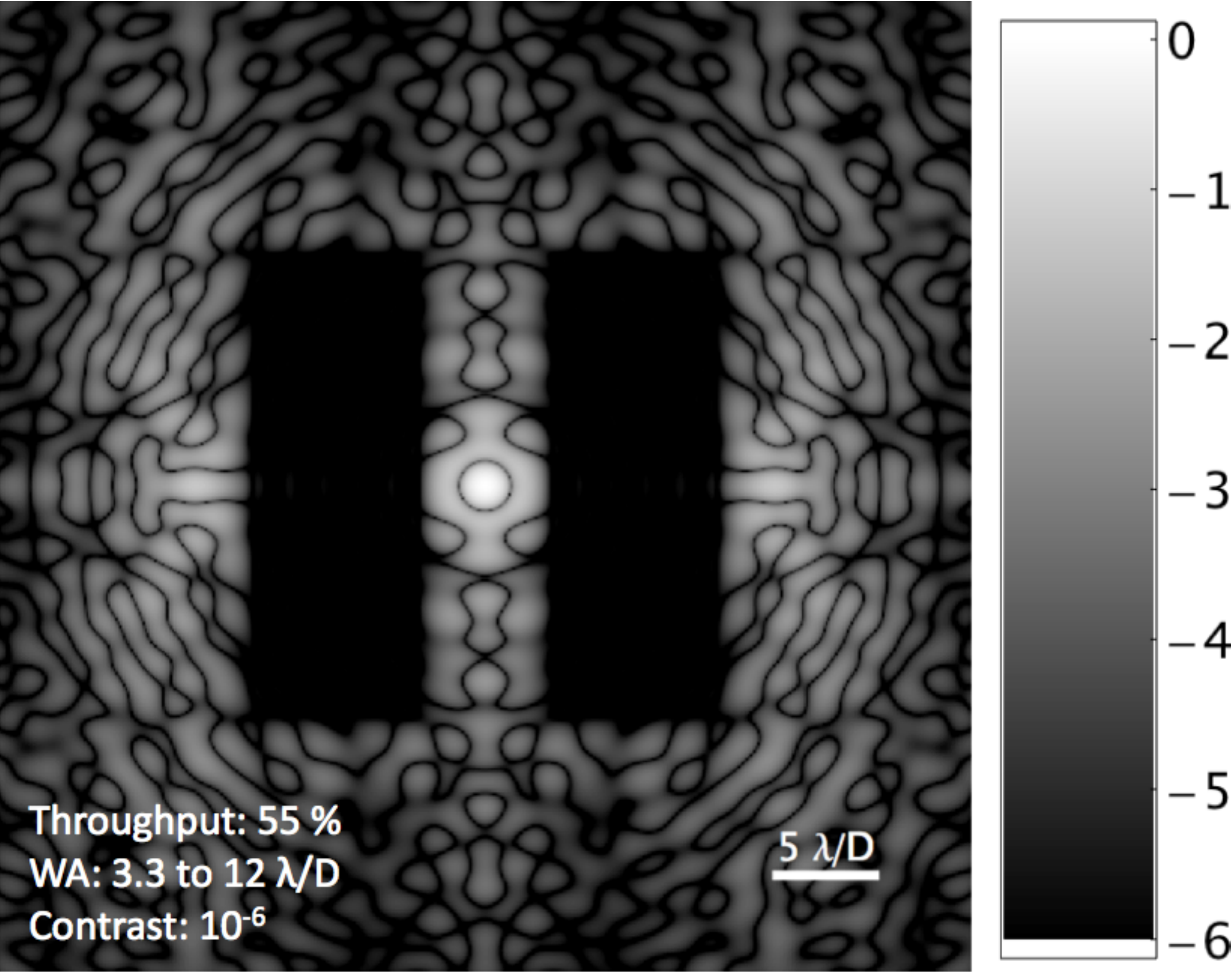}\\
\end{tabular}
\end{center}
\caption{Pupil mask (left) designed to reach $10^{-6}$ with the SPICA telescope. The high contrast areas can be seen on the  associated PSF (right). The IWA is 3.5 $\lambda/D$ and the OWA is 12 $\lambda/D$. The vertical width of the dark hole is 50 $\lambda/D$.}
\label{SPICA}
\end{figure}

Several masks for SPICA, as well as for Subaru and JWST, have been described in \cite{Enya2010, Enya2011}.
Different strategies have been explored that favor the contrast over the IWA or vice-versa.
The mask we present here, shown in Figure~\ref{SPICA}, has a throughput of 55\%. Its IWA is 3.3 $\lambda/D$ and its OWA is 12 $\lambda/D$. The discovery zone for which it is designed is similar to the one created by the first mask that appears in \cite{Enya2011}.
Because of the absence of any constraints on the optimization of the mask, a significant gain in throughput is  obtained (the light is transmitted in an additional one fourth of the pupil surface).

\subsection{Subaru}

Unlike SPICA, the spiders of Subaru are not perpendicular. A solution proposed in \cite{Enya2010} was to place two barcode masks in a direction perpendicular to one of the spiders so that the design of the mask can be affected as little as possible. This way the mask can provide a contrast of $10^{-5}$ at 3 $\lambda/D$ for a throughput of 24\%.

\begin{figure}
\begin{center}
\begin{tabular}{ccl}
\includegraphics[height=5cm]{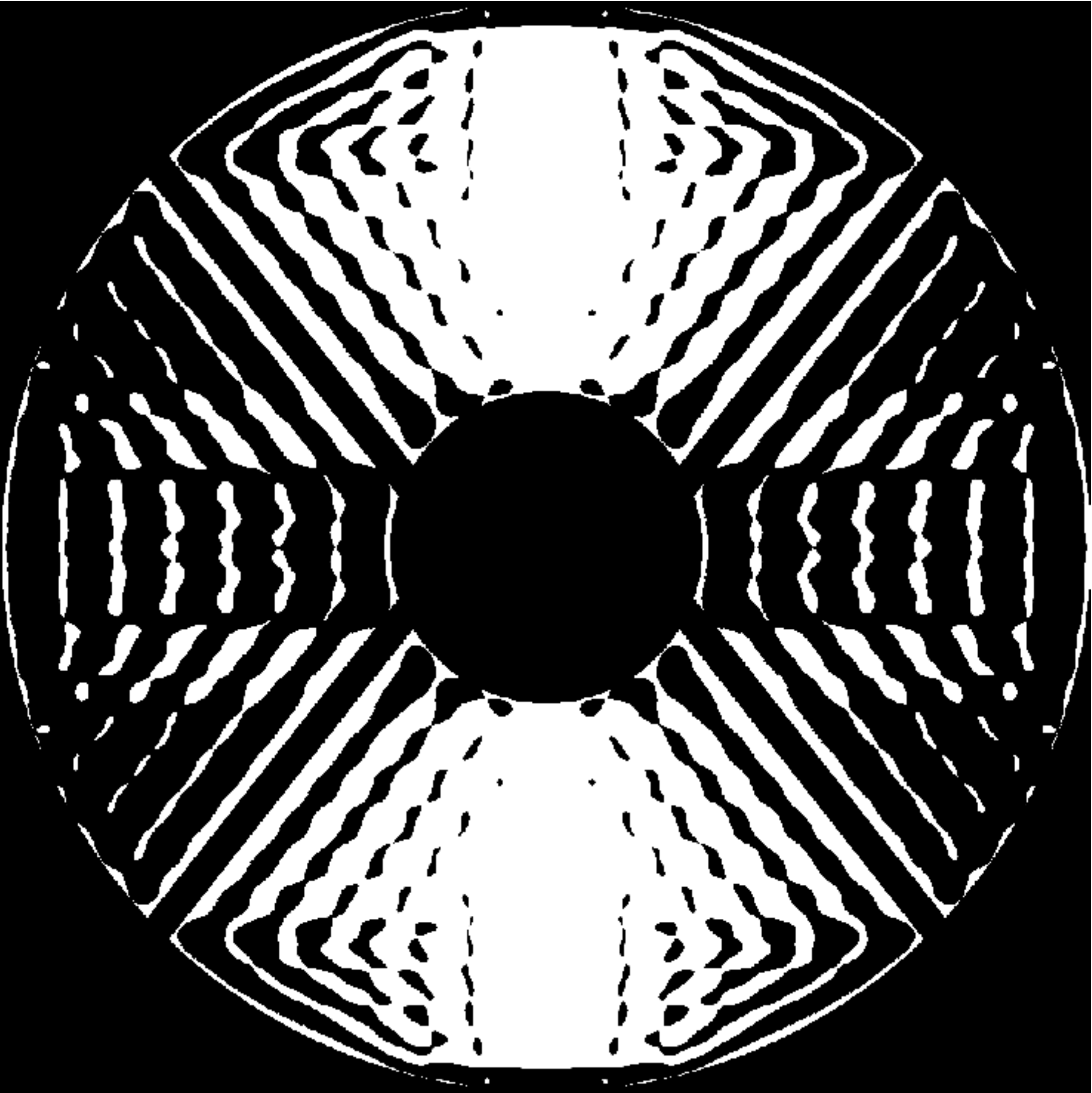} & \includegraphics[height=5cm]{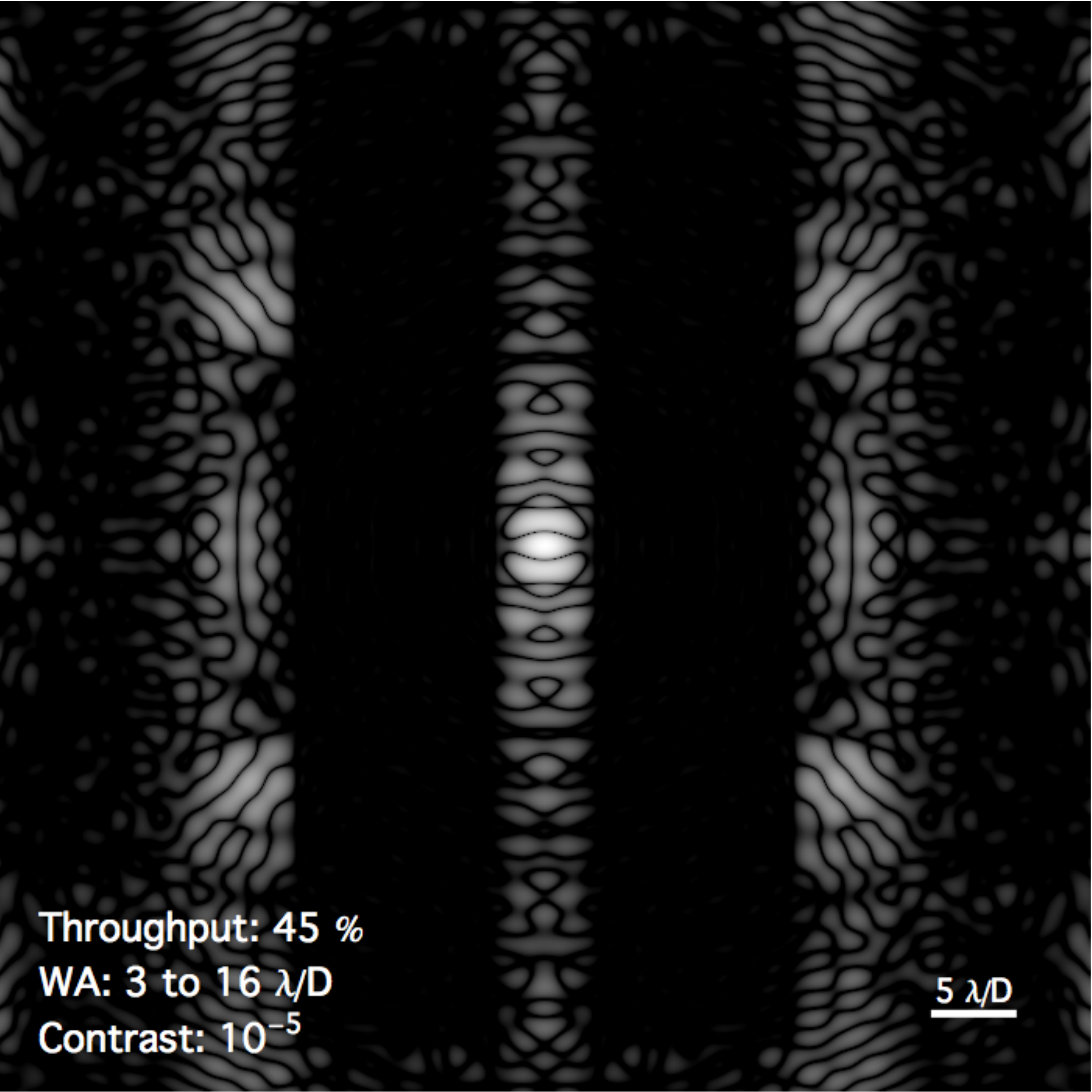} & \includegraphics[width=1.04cm]{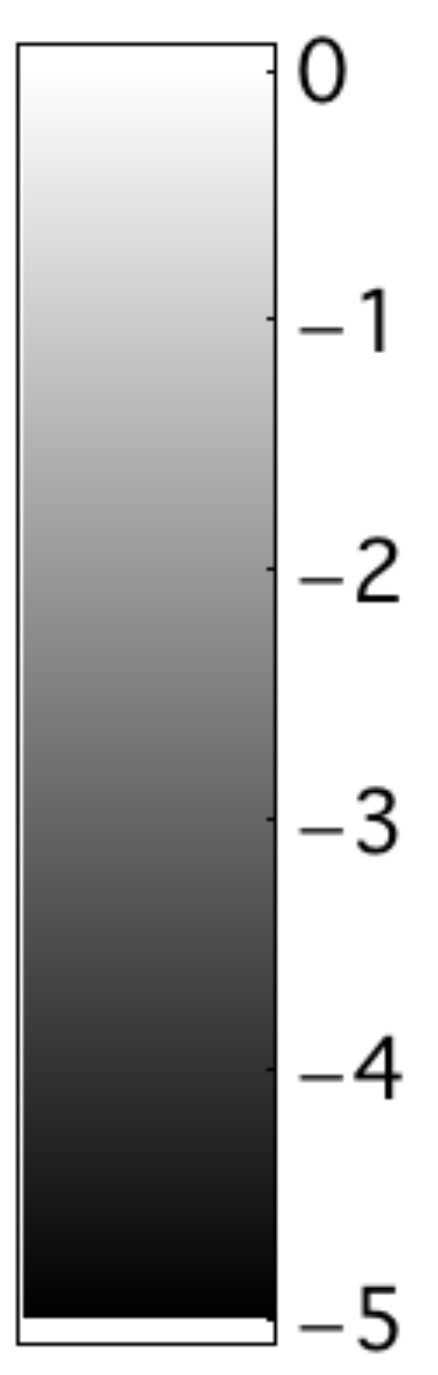}\\
\includegraphics[height=5cm]{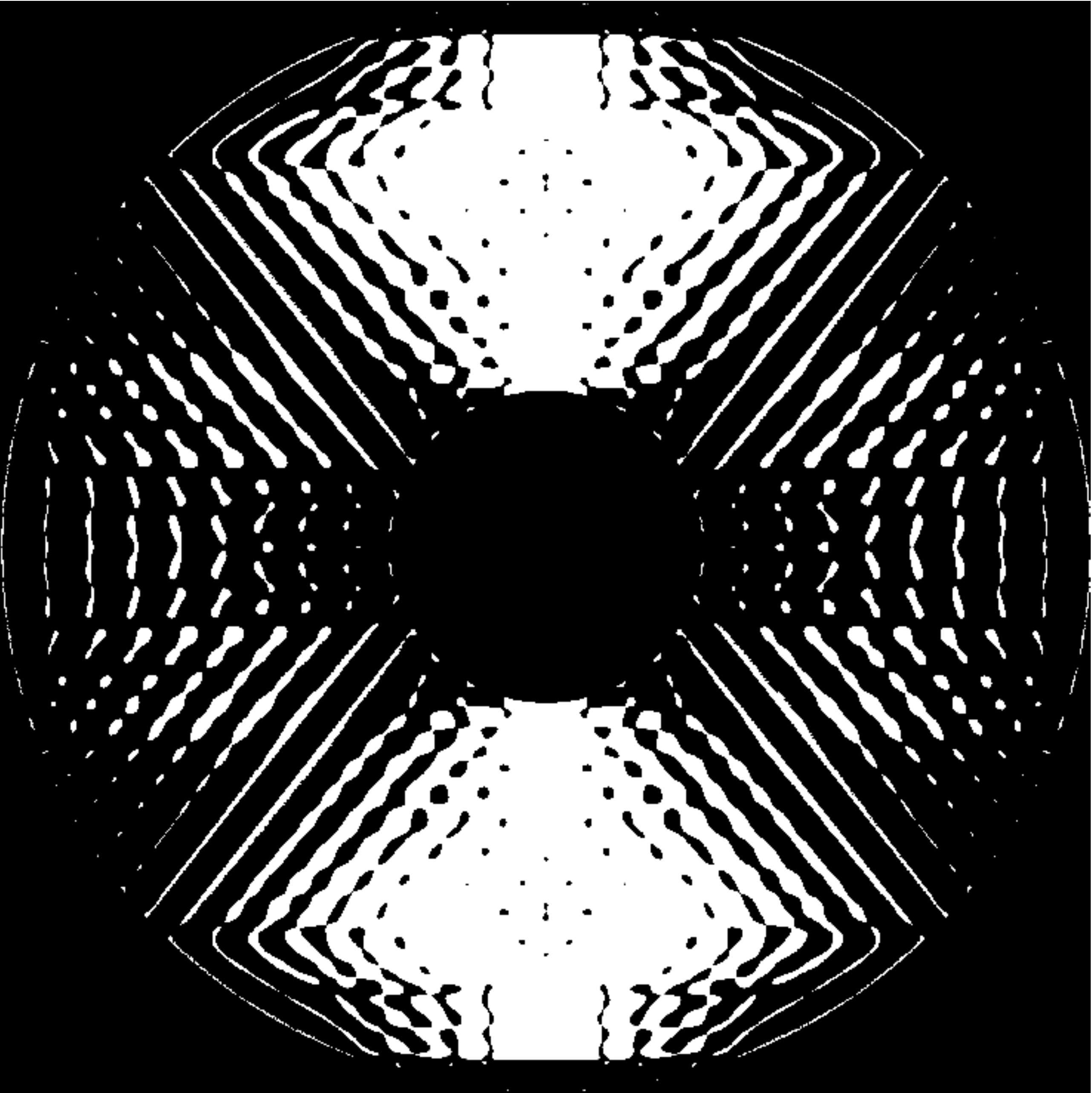} & \includegraphics[height=5cm]{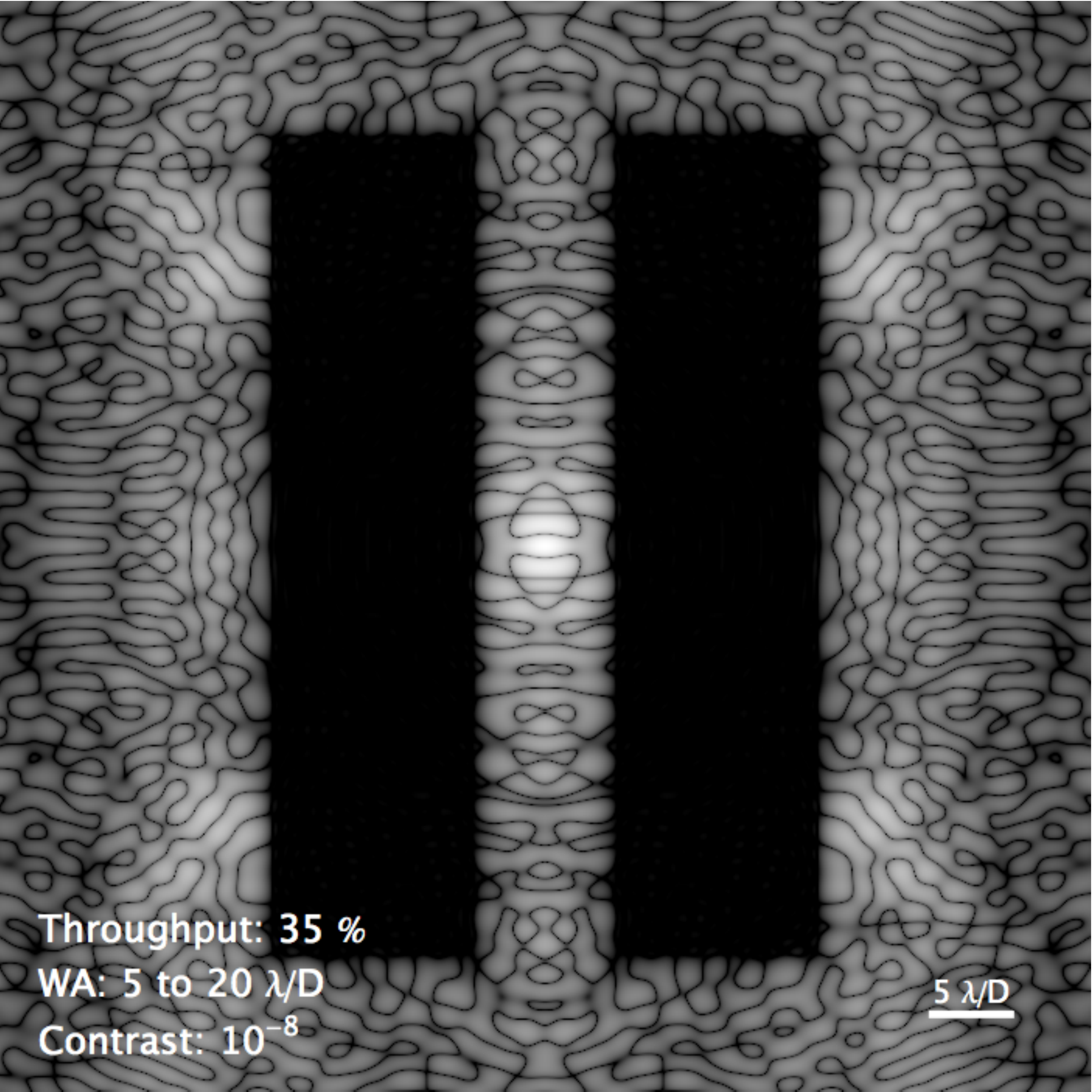} & \includegraphics[width=0.9cm]{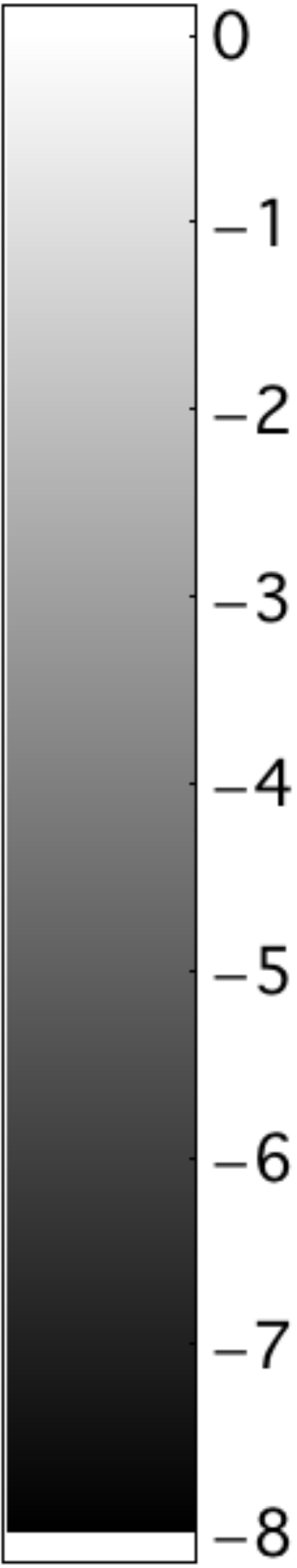}
\end{tabular}
\end{center}
\caption{Pupil masks (left) designed to reach $10^{-5}$ (top) and $10^{-8}$ (bottom) with the Subaru telescope. The high contrast areas can be seen on their  associated PSFs (right). The IWA are 3 and 5 $\lambda/D$, the OWA are 16 and 20 $\lambda/D$, and the vertical widths of the dark holes are 64 and 60 $\lambda/D$.}
\label{Subaru}
\end{figure}

The first mask we present here was designed with the same constraints as Mask-1 in \cite{Enya2010} but rotated by $\approx 38$ degree. The particular orientation chosen in their case is necessary so as to display the barcode masks horizontally. We are not subject to the same requirement, and we have chosen to align the axes of symmetry of the central obstruction with the x and y-axes of the simulation window. The top of Figure~\ref{Subaru} displays the mask and its PSF. It allows through almost twice the light, with a throughput of 45\%. 

The second mask was designed to reach a higher contrast, $10^{-8}$, although a larger IWA of 5 $\lambda/D$ is necessary for the throughput to remain comparable (35\% of the light is transmitted).

\section{Asymmetric pupil with a segmented mirror}
\label{sec4}

We now apply the mask optimization to  the James Webb Space Telescope (JWST) pupil. The pupil mask is optimized with N=1024. Contrary to all the preceding pupils we have considered, this pupil only has one axis of symmetry. Because of that we cannot optimize it in the exact same way as we did for the others. We still take advantage of one axis of symmetry in the computation of the first Fourier transform (and one should make sure that the pupil has the correct orientation, or rotate it by $\pi/2$ if necessary), but the second FT must be completed without it. This results in an array twice as large as with the other pupils ($512 \times 1024$).

Another consequence of the asymmetry is that the exponential term in the expression of the FT cannot be changed into a cosine.  Hence each number has a real and an imaginary part, which doubles the amount of memory required to solve this model.  Furthermore, to keep the problem linear, the natural contrast constraint $|E(u,v)| \le 10^{-2.5} E(0,0)$ is replaced with a pair of constraints, one for the real part and one for the imaginary part:
\begin{eqnarray*}
    -10^{-2.5} E(0,0)/\sqrt{2} \quad \le & \Re(E)(u,v) & \le \quad 10^{-2.5} E(0,0)/\sqrt{2} \\
    -10^{-2.5} E(0,0)/\sqrt{2} \quad \le & \Im(E)(u,v) & \le \quad 10^{-2.5} E(0,0)/\sqrt{2} .
\end{eqnarray*}

\begin{figure}
\begin{center}
\begin{tabular}{ccl}
\includegraphics[height=5cm]{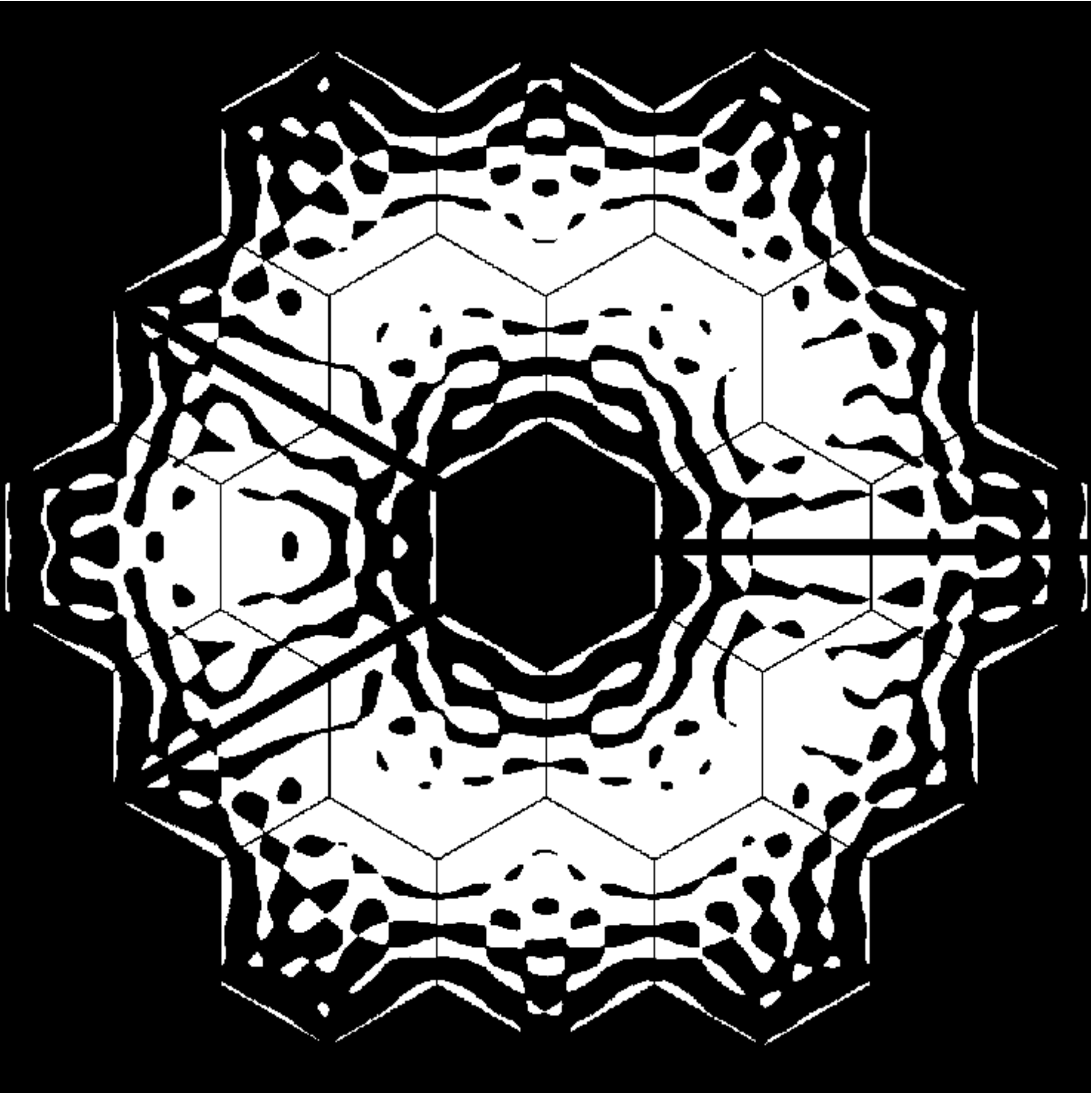} & \includegraphics[height=5cm]{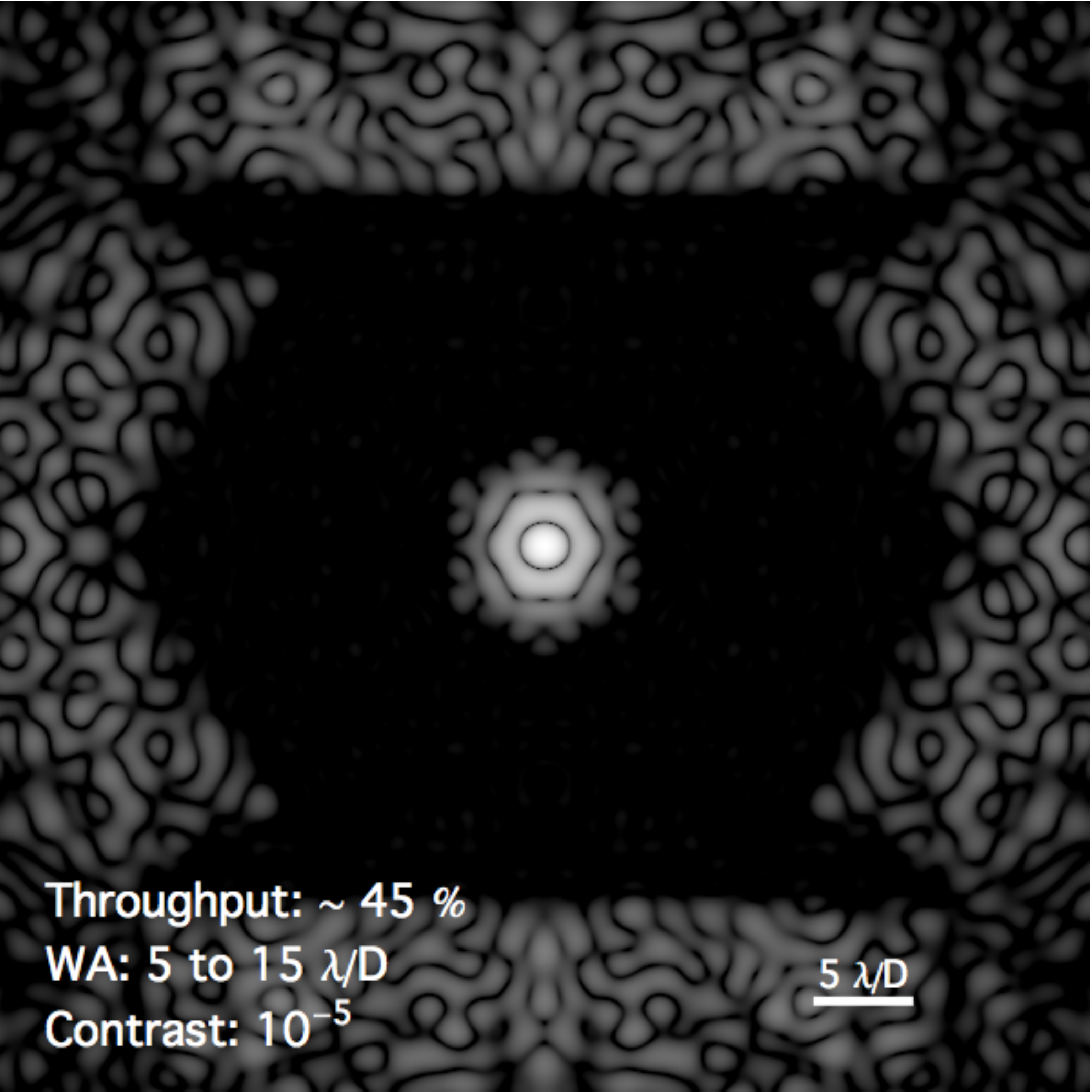} & \includegraphics[width=1.04cm]{ContrastBar5.pdf}\\
\end{tabular}
\end{center}
\caption{Pupil mask (left) designed to reach $10^{-5}$ with the JWST. The hexagonal high contrast area can be seen on its  associated PSF (right). The IWA and OWA are 5 and 15 $\lambda/D$.}
\label{JWST}
\end{figure}

The JWST pupil is particularly interesting: the internal and external hexagonal segments making up the pupil have a great effect on the contrast that can be reached. A first mask was generated for a circular ring dark hole with a minimal IWA of 5 $\lambda/D$. A smaller IWA (4 $\lambda/D$) could be obtained with a similar throughput by changing the dark hole to a hexagonal ring, with internal and external contours rotated by $\pi/2$  with respect to each other. The resulting mask and PSF is shown in Figure~\ref{JWST}.  Several other geometric aspects still remain to be tested. 

\section{Final Remarks}
\label{sec5}

Previous two-dimensional shaped-pupil optimizations all assumed specific geometries in order to reformulate the 2D problem as a 1D mathematical description that could then be solved. We have shown in this paper that full optimizations of pupil masks in two dimensions are possible. No constraint other than the initial geometry of the aperture are placed on the pupil.  This allows us to find new, and more optimal, solutions to a variety of high contrast problems. We have in particular compared composite barcode masks designed for the SPICA coronagraph and for the Subaru telescope to new, fully optimized 2D masks. Given the same discovery zones (same contrasts, IWA and OWA), higher throughputs can always be expected. More importantly, these throughputs are simply the highest possible.

Complex pupil geometries can be treated without much more efforts than basic geometries. Masks optimized for the JWST illustrate particularly well the possibilities that full 2D optimizations offer. Despite the central obstruction, the spiders and the inter-segments gaps, optimal mask solutions are found.

Another very important observation is that the result of the optimization always tends to be a binary mask, and that, given a sufficiently high enough number of points used to sample the pupil plane, it virtually is. This had already been observed when optimizing 1D masks, but the optimization of ripple masks and star shaped masks assumed a binary amplitude distribution.

One can argue that these optimizations are constrained by the choice of the high contrast regions. In the basic case of a circular symmetric pupil, without central obstruction and spiders, many types of dark holes can be considered (circular symmetric rings, portions of rings, rectangular areas, etc.), but the particularities of pupils with more complex features must be taken into account when designing the dark holes.  For example, the small size of spiders leads to diffraction effects on a large scale, and the geometry of the spiders must be used to design the dark holes. A system of 4 spiders set in an orthogonal fashion will create diffraction spikes in a similar orthogonal fashion. One can then choose to move the diffracted energy along only one axis, so as to obtain a good contrast on the other, or one can also choose to maximize the contrast in the 4 quadrants of the image plane by concentrating the diffracted energy along both axes.

Many coronagraphs\footnote{in particular the PIAA, the FQPM, and the vortex coronagraph} are meant to be used with telescopes without central obstructions. Optical devices can be used to artificially remove the spiders and/or the central obstruction. Masks could be designed, not to aim for high contrasts, but to attenuate the diffraction spikes. The advantage of choosing this solution would be a better achromaticity and little to no additional aberrations introduced in the wavefront. This will be explored in a future paper.


The masks that are presented here are all binary, but none of them is fully structurally connected, in contrast to bar code masks, star shaped masks, or shaped pupils. This makes it mandatory for them to be laid on glass or mirrors. Additional amplitude and phase aberrations will thus be introduced on the wavefront, and will have to be taken into account in the design of the wavefront correction system. Checkerboard masks are also not structurally connected, and some were manufactured for the SPICA project using e-beam lithography of aluminum coated BK7 glass substrates. Masks designed for $10^{-10}$ were tested in the laboratory, and contrasts smaller than $7 \times 10^{-8}$ were achieved in laser light \cite{Enya2011}. More recent experiments were conducted in broadband, and in a vacuum chamber (see \cite{Haze2011}). This time the contrast was limited by ghosts and ranged from $3.5 \times 10^{-7}$ for a central wavelength $\lambda_{0}$ = 650 nm ($\Delta\lambda$ = 8 nm) to $2.6 \times 10^{-6}$ for $\lambda_{0}$ = 850 nm ($\Delta\lambda$ = 55 nm). In both experiments the wavefront was not actively corrected. 

We intend in the near future to have some masks manufactured in a similar fashion, and use them in our high contrast imaging laboratory in Princeton, together with our 2 DM wavefront correction system.


\end{document}